\pdfoutput=1
\PassOptionsToPackage{unicode}{hyperref}
\PassOptionsToPackage{hyphens}{url}
\PassOptionsToPackage{dvipsnames,svgnames,x11names}{xcolor}
\documentclass[
  10pt,
]{article}
\usepackage{amsmath,amssymb}
\usepackage{iftex}
\ifPDFTeX
  \usepackage[T1]{fontenc}
  \usepackage[utf8]{inputenc}
  \usepackage{textcomp} 
\else 
  \usepackage{unicode-math} 
  \defaultfontfeatures{Scale=MatchLowercase}
  \defaultfontfeatures[\rmfamily]{Ligatures=TeX,Scale=1}
\fi
\usepackage{lmodern}
\usepackage[english]{babel}
\ifPDFTeX\else
\fi
\IfFileExists{upquote.sty}{\usepackage{upquote}}{}
\IfFileExists{microtype.sty}{
  \usepackage[]{microtype}
  \UseMicrotypeSet[protrusion]{basicmath} 
}{}
\makeatletter
\@ifundefined{KOMAClassName}{
  \IfFileExists{parskip.sty}{%
    \usepackage{parskip}
  }{
    \setlength{\parindent}{0pt}
    \setlength{\parskip}{6pt plus 2pt minus 1pt}}
}{
  \KOMAoptions{parskip=half}}
\makeatother
\usepackage{xcolor}
\usepackage[margin=2.3cm]{geometry}
\usepackage{longtable,booktabs,array}
\usepackage{calc} 
\usepackage{etoolbox}
\makeatletter
\patchcmd\longtable{\par}{\if@noskipsec\mbox{}\fi\par}{}{}
\makeatother
\IfFileExists{footnotehyper.sty}{\usepackage{footnotehyper}}{\usepackage{footnote}}
\makesavenoteenv{longtable}
\providecommand{\tightlist}{%
  \setlength{\itemsep}{0pt}\setlength{\parskip}{0pt}}
\usepackage{amsmath,amssymb}
\usepackage{graphicx}
\usepackage{cite}                 
\usepackage{caption}
\usepackage{etoolbox}
\usepackage{framed}
\definecolor{shadecolor}{gray}{0.95}
\AtBeginEnvironment{longtable}{\small}
\DeclareRobustCommand{\_}{\textunderscore\hspace{0pt}}
\DeclareUnicodeCharacter{2212}{\ensuremath{-}}
\DeclareUnicodeCharacter{2192}{\ensuremath{\rightarrow}}
\DeclareUnicodeCharacter{03C1}{\ensuremath{\rho}}
\DeclareUnicodeCharacter{03B2}{\ensuremath{\beta}}
\DeclareUnicodeCharacter{0394}{\ensuremath{\Delta}}
\DeclareUnicodeCharacter{03C3}{\ensuremath{\sigma}}
\DeclareUnicodeCharacter{03BA}{\ensuremath{\kappa}}
\DeclareUnicodeCharacter{2265}{\ensuremath{\geq}}
\DeclareUnicodeCharacter{2264}{\ensuremath{\leq}}
\DeclareUnicodeCharacter{2248}{\ensuremath{\approx}}
\DeclareUnicodeCharacter{00B3}{\ensuremath{{}^{3}}}
\DeclareUnicodeCharacter{2011}{\mbox{-}}
\DeclareUnicodeCharacter{202F}{\,}
\ifLuaTeX
  \usepackage{selnolig}  
\fi
\IfFileExists{bookmark.sty}{\usepackage{bookmark}}{\usepackage{hyperref}}
\IfFileExists{xurl.sty}{\usepackage{xurl}}{} 
\hypersetup{
  pdftitle={Do in-silico metrics filter designed enzymes for catalysis?},
  pdfauthor={Ziqi Zhang; Wei Zhang; Zhaohong Deng},
  colorlinks=true,
  linkcolor={blue},
  filecolor={Maroon},
  citecolor={Blue},
  urlcolor={blue},
  pdfcreator={LaTeX via pandoc}}

\newcommand{\hc}[1]{\begin{minipage}[b]{\linewidth}\raggedright #1\end{minipage}}

\title{Do in-silico metrics filter designed enzymes for catalysis?}
\author{%
  Ziqi Zhang\textsuperscript{1}, Wei Zhang\textsuperscript{2}, Zhaohong Deng\textsuperscript{1,*}\\[6pt]
  {\small\textsuperscript{1}School of Artificial Intelligence and Computer Science, Jiangnan University, Wuxi, Jiangsu 214122, China}\\
  {\small\textsuperscript{2}School of Artificial Intelligence and Computer Science, Nantong University, Nantong, Jiangsu 226019, China}\\[4pt]
  {\small\textsuperscript{*}Corresponding author: \texttt{dengzhaohong@jiangnan.edu.cn}}%
}
\date{}

\begin{document}
\maketitle

\nocite{listov2026,
  butcher2025,
  kim2026,
  listov2025,
  richter2011,
  abramson2024,
  lauko2025,
  ahern2026,
  watson2023,
  yeh2023,
  braun2026,
  anishchenko2025,
  chai2024,
  wohlwend2024,
  passaro2025,
  krishna2024,
  jumper2021,
  lin2023,
  zhang2025sefp,
  zhang2025phopt,
  zhang2025segment,
  zhang2026patchet,
  shen2025asceppis,
  zhang2025m2st,
  zhang2026ssmgcn,
  johnson2025,
  overath2025,
  garcia2026,
  korbeld2026,
  kiss2010,
  alford2017,
  chaudhury2010,
  lee1971,
  mitternacht2016,
  tien2013,
  olsson2011,
  sondergaard2011,
  evans2021,
  burley2025,
  berman2000,
  uniprot2023,
  dana2019,
  ribeiro2018,
  hanley1982,
  mann1947,
  efron1979,
  field2007,
  westfall1993,
  carlin2016,
  carlin2017,
  spearman1904}

\section*{Abstract} \addcontentsline{toc}{section}{Abstract}

Computational metrics are widely used to prioritize enzyme designs for experimental testing, yet their ability to capture catalytic function remains poorly established. Here, we systematically evaluate 33 in-silico metrics across four complementary tasks: distinguishing active from inactive enzymes, identifying substrate specificity, ranking enzyme variants by activity, and detecting perturbations to catalytic residues. We assess these capabilities using natural enzymes, computationally designed enzymes, activity-altering variants, and controlled structural perturbations that compare catalytic residues with matched non-catalytic sites. No single metric consistently captures all aspects of catalytic function. In activity discrimination, catalytic-residue intra-residue repulsion achieves the highest AUROC among the evaluated metrics for both natural enzymes (0.61) and 192 designs that passed the filters of their original design campaign (0.80). For substrate discrimination, interface predicted TM-score (ipTM) performs best (AUROC 0.70), whereas the catalytic pKa range shows the strongest association with activity across single-point variants of BglB (Spearman’s ρ = 0.48). In controlled perturbation tests, the minimum per-chain predicted TM-score (pTM) has a mean probability of 0.75 of responding more strongly to perturbations at catalytic residues than to matched perturbations at non-catalytic sites. These results reveal substantial variation in what current computational metrics capture, underscoring the need to distinguish structural confidence and physical plausibility from direct evidence of catalytic function when prioritizing enzyme designs.

\section{1 Introduction}

Computational enzyme design \cite{listov2026} now produces many more candidates than can be tested.
A single design problem yields from more than a hundred thousand to millions of designs: 130,000
backbones for a cysteine hydrolase designed with RFdiffusion3 \cite{butcher2025}, several hundred
thousand sequences for a zinc metallohydrolase \cite{kim2026}, and 100,000 to 1,000,000 designs for
each starting structure in a Kemp eliminase workflow \cite{listov2025}. The experiment, one or two
96-well plates, has room for one or two hundred designs, and an in-silico filter decides which.
Richter and colleagues described this filtering step in 2011 as ``analyzing, and ranking all of the
produced structures to find the best handful worth experimentally characterizing''
\cite{richter2011}.

Recent campaigns have invested heavily in this step. The cysteine hydrolases of RFdiffusion3 went
through three rounds of filtering based on AlphaFold3 \cite{abramson2024} predictions of the
catalytic geometry, and 18\% of the 190 designs tested were multi-turnover enzymes, against 1--2\%
in earlier campaigns \cite{butcher2025}. In the serine hydrolase campaign, the fraction of designs
with ester hydrolysis activity rose from 1.6\% to 5.2\% in the round in which a filter for the
preorganisation of the catalytic residues was added \cite{lauko2025}. In-silico criteria also decide
which method is judged to be better: RFdiffusion2 delivers a design that passes them for all 41
active sites of its benchmark, RFdiffusion for 16 \cite{ahern2026,watson2023}. Filtering is thus a
decisive step in enzyme design, but passing it does not guarantee activity: of the 192 designs
tested in the metallohydrolase campaign, 176 have no reported kcat/KM \cite{kim2026}.

The filters themselves are far from standardised: what they measure differs from pipeline to
pipeline. Some rank designs on the basis of Rosetta energies and shape complementarity
\cite{kim2026,listov2025,yeh2023}, others on how accurately a predicted structure reproduces the
intended catalytic geometry \cite{butcher2025,braun2026}, still others on the preorganisation of the
active site in an ensemble of predicted conformations \cite{lauko2025,anishchenko2025}, and yet
others on the confidence of AlphaFold 3 \cite{abramson2024} or Chai-1 \cite{chai2024} models of the
enzyme--substrate complex. Prediction tools can serve as a filter
\cite{abramson2024,anishchenko2025,chai2024,wohlwend2024,passaro2025,krishna2024,jumper2021,lin2023,zhang2025sefp,zhang2025phopt,zhang2025segment,zhang2026patchet,shen2025asceppis,zhang2025m2st,zhang2026ssmgcn}.
Many pipelines combine several of them into a composite score
\cite{butcher2025,kim2026,listov2025}. The cut-offs differ from campaign to campaign, and the
weights are set at the designers' discretion, in one case as ``an arbitrary first guess''
\cite{kim2026}.

That judgement may rest on little evidence: it is not known whether any of these metrics responds to
catalysis. A campaign assesses its filter by its own hit rate, which also reflects every other
change between rounds. For the final selection, the designers of the metallohydrolases combined many
metrics instead of applying cut-offs ``because of the lack of known standalone predictive metrics
for metallohydrolase activity'' \cite{kim2026}. The authors of Riff-Diff, who ranked their designs
mainly on the RMSD of the catalytic residues, report a ``negligible correlation between predicted
enzyme rankings and actual activity in this and previous studies'' \cite{braun2026}. For adjacent
problems, systematic evaluations exist: filters for generated enzyme sequences \cite{johnson2025},
interface metrics for protein binders \cite{overath2025}, predictor confidence for monomer designs
\cite{garcia2026,korbeld2026} and, in 2010, quantum-mechanical and molecular-dynamics scoring of the
first Kemp eliminase designs \cite{kiss2010}. To our knowledge, none of them has examined whether
the fast metrics used to filter structure-based enzyme designs respond to catalysis.

To address this, we ask what a metric that tracks catalysis should be able to do, and test each
answer. It should be able to distinguish an enzyme that works from an enzyme that does not, which is
what a filter is for (\emph{activity discrimination test}). It should also know which substrate an
enzyme was built for, as every design conditioned on a substrate presupposes (\emph{substrate
discrimination test}), and it should order enzymes by their activity, because a selection ends with
choosing the best of the designs that pass (\emph{activity ranking test}). And it should notice
damage to the catalytic residues more strongly than the same damage elsewhere in the protein, since
that is what distinguishes catalysis from geometry (\emph{detection ability test}). We applied the
tests to 33 metrics that can be computed from a design model or a co-folding prediction (Methods
2.1). The first three look at natural enzymes and designs as they are, whereas the fourth
deliberately damages natural enzymes and compares the response at the catalytic residues with the
response at a matched control.

In activity discrimination on natural enzymes, the intra-residue repulsion of the catalytic residues
ranked first of the 33 metrics (AUROC 0.61), against a chance threshold of 0.68 for the best of 39
quantities. On the 192 designs, which had passed the filters of their campaign, intra-residue
repulsion ranked first (0.80) and the repulsion of the catalytic residues second (0.78). In
substrate discrimination, ipTM ranked first (0.70) and the RMSD of the catalytic atoms reached 0.53
to 0.56. In activity ranking, the catalytic pKa range ranked first among the single-point variants
of the enzyme BglB (Spearman ρ = 0.48). In detection ability, the per-chain pTM minimum had the
highest mean R over the counted steps (0.75; 0.5 means no preference for the catalytic site)
(Results 3.1 to 3.4).

\begin{center}\rule{0.5\linewidth}{0.5pt}\end{center}

\section{2 Methods}

\subsection{2.1 Overview of the experiment}

Four tests, one in each of 2.2 to 2.5, examine whether the in-silico metrics used to filter
designed enzymes respond to catalysis:

\begin{itemize}
\tightlist
\item \textbf{Activity discrimination test (2.2; results in 3.1).} The metrics are ranked by how
well they actually separate dead from active proteins: 21 zymogen--mature enzyme pairs and 192
plated de novo designs, of which 16 are active.
\item \textbf{Substrate discrimination test (2.3; results in 3.2).} A structure predictor is given
each of 55 natural enzymes with its own substrate and with a wrong ligand, and the prediction-based
metrics are ranked by how well they separate the two predictions.
\item \textbf{Activity ranking test (2.4; results in 3.3).} The metrics are ranked by how well their
order of 432 BglB variants and of the 16 plated designs with a measured activity agrees with the
measured activity.
\item \textbf{Detection ability test (2.5; results in 3.4).} Catalytic residues are deliberately
damaged in up to 195 natural enzymes, and the metrics are ranked by whether they change more than
when the same damage is applied at a matched non-catalytic site of the same protein.
\end{itemize}

Figure 1 follows each dataset from its source to the evaluation set on which each table of the
Results is computed; the same sets are described in 2.2 to 2.5.

\textbf{Metrics.} The same panel is scored in every test: 29 structure-space metrics, which read
atomic coordinates and are used in 2.2, 2.4 and 2.5, and 4 prediction-based metrics, which read the
output of a structure predictor that is given a sequence and a ligand and are used in 2.2 to 2.5
(Table 1). Numbers in square brackets are references (see References); tables with an S-number are
supplementary tables, described in the Supplement PDF and released as CSV files (\texttt{main/} and
\texttt{supplementary/}).

\begin{longtable}[]{@{}
  >{\raggedright\arraybackslash}p{(\columnwidth - 6\tabcolsep) * \real{0.1613}}
  >{\raggedright\arraybackslash}p{(\columnwidth - 6\tabcolsep) * \real{0.3226}}
  >{\raggedright\arraybackslash}p{(\columnwidth - 6\tabcolsep) * \real{0.3548}}
  >{\raggedright\arraybackslash}p{(\columnwidth - 6\tabcolsep) * \real{0.1613}}@{}}
\caption*{\textbf{Table 1. The 33 metrics.} In a structure, ``perturbed residues'' are the residues
that have been changed; in the control arm they are the control residues, so a metric scored per
site never knows which residues are catalytic --- it only knows which residues it has been given.
In 2.2 and 2.4 the scored residues are the catalytic residues. The third column shows where each
definition comes from.}\tabularnewline
\toprule\noalign{}
\hc{Group} & \hc{Metrics} & \hc{Reads, and where it is defined} & \hc{Scored over} \\
\midrule\noalign{}
\endfirsthead
\toprule\noalign{}
\hc{Group} & \hc{Metrics} & \hc{Reads, and where it is defined} & \hc{Scored over} \\
\midrule\noalign{}
\endhead
\bottomrule\noalign{}
\endlastfoot
Rosetta energy, scored per site (15) &
attraction, repulsion, intra-residue repulsion, solvation, lk-ball solvation, electrostatics,
hydrogen bonds (backbone--side chain, side chain--side chain), rotamer (Dunbrack), Ramachandran,
amino-acid propensity, omega torsion, net van der Waals (repulsion plus attraction), site total,
worst single residue &
coordinates; the per-residue terms of the REF2015 energy function \cite{alford2017}, computed with
PyRosetta \cite{chaudhury2010} and summed over the scored residues (site total: the sum of the
residue totals; worst residue: the largest residue total) & the perturbed residues \\
Catalytic geometry (8) & fraction buried, relative solvent-accessible surface area (mean, max),
total solvent-accessible surface area, pairwise distance (mean, max), radius of gyration, polar
contacts &
coordinates; SASA by the Lee--Richards algorithm \cite{lee1971} in FreeSASA
\cite{mitternacht2016} (probe 1.4 Å, water excluded), relative to the maximum accessibility of the
residue type \cite{tien2013}; the fraction buried is the share of the scored residues with a
relative SASA below 0.10; distances and radius of gyration are computed over all pairs, respectively
all atoms, of the heavy side-chain atoms of the scored residues (backbone atoms for Gly); a polar
contact is a pair of polar side-chain atoms of two scored residues within 3.5 Å & the perturbed
residues \\
Catalytic pKa (6) &
PROPKA pKa (min, max, mean, range) and pKa shift (mean, maximum of the absolute value) &
coordinates; the pKa from PROPKA3 \cite{olsson2011,sondergaard2011} for the scored ionisable
residues; the shift is the pKa minus the pKa of the model compound & the perturbed residues \\
Interface confidence (2) & ipTM; per-chain pTM minimum & output of the Chai-1 predictor
\cite{chai2024}; ipTM as introduced for AlphaFold-Multimer \cite{evans2021}; the pTM of each chain,
from the TM-score head of AlphaFold \cite{jumper2021} &
the protein--ligand interface \\
Active-site accuracy (2) & AME RMSD (RMSD of the catalytic atoms relative to a reference structure
after backbone superposition, best of the predicted models); ligand clearance (closest approach of
the ligand to the backbone, best of the models) &
output of the predictor and a reference structure; adapted from the criteria of the Atomic Motif
Enzyme benchmark of RFdiffusion2 \cite{ahern2026} (comparisons in 2.3) &
the catalytic atoms; the ligand \\
\end{longtable}

\textbf{The metric panel.} The software returns 155 named quantities, which amount to 139 distinct
metrics once aliases and the per-model summaries of a single score (mean, maximum, spread) are
merged; 61 of them are bookkeeping counters and constants. Table 1 lists the 33 metrics that are
valid in all four tests. The combined Chai-1 score is a rescaled ipTM and not an independent metric;
its AUROC values are given in 2.3. Table S1 gives the class of each of the 139 distinct metrics, and
Table S2 what each of the 155 named quantities reads and over which residues it is scored.

\textbf{Notation.} Every table in the Results ranks a set of \emph{items} (the metrics): \(m\) is
an item, \(M\) the number of quantities analysed in the test of one table, which may exceed the
number of items listed in it, and \(x_u(m)\) the value of item \(m\) on a unit \(u\) (a structure, a
prediction, a variant or a design). Bootstrap and permutation tests use \(N_{\mathrm{boot}} =
N_{\mathrm{perm}} = 5{,}000\) draws unless another number is stated; \(q_{\alpha}\) is the
\(\alpha\)-quantile of a set of values and \(\mathbf{1}(\cdot)\) is 1 if the argument is true and 0
otherwise. The statistics (\(A_m\), \(\rho_m\), \(R_{m,s}\) and the best-of-\(M\) threshold from the
max statistic \(T^{(r)}\)) are defined with their equations in 2.2 to 2.5. The best-of-\(M\)
thresholds are reference levels for reading a ranking, and each table states the evaluation set on
which it was computed.

\begin{figure}[!t]
\centering
\includegraphics[width=\linewidth]{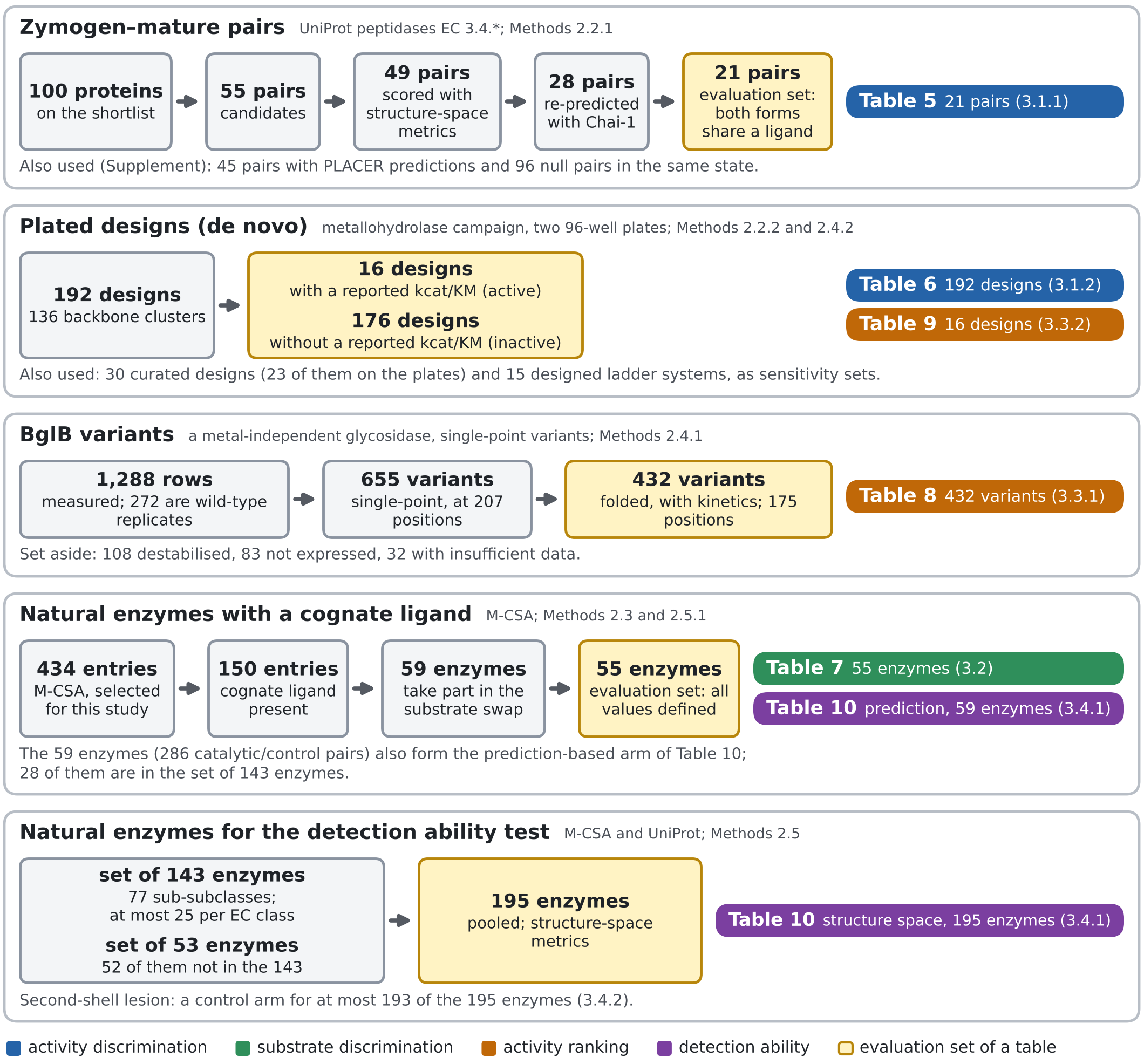}
\par\smallskip

{\small

\textbf{Figure 1. Datasets and evaluation sets.} Each dataset from its source to the evaluation set
(shaded) on which a table of the Results is computed, with the number of cases at each step; the
tables of the Results are coloured by test. Further sets described in the Results are in the
Supplement.

}

\end{figure}

\begin{center}\rule{0.5\linewidth}{0.5pt}\end{center}

\subsection{2.2 Activity discrimination test: do the metrics separate a dead enzyme from an active
enzyme?}

\textbf{Experiment.} Each metric is scored on real proteins in which one state is catalytically
dead and the other active, and the metrics are ranked by how well their values separate the two
states. A zymogen, for example, is an inactive precursor that already contains the catalytic
residues of the mature enzyme: a metric that detects catalytic competence would separate the two
structures, whereas a metric that reads only the geometry of the catalytic residues would not,
because that geometry is almost the same in both forms. Two datasets are used, each with its own
table and its own n: 21 zymogen--mature pairs (2.2.1; results in 3.1.1, Table 5) and 192 plated
designs, of which 16 are active (2.2.2; results in 3.1.2, Table 6). The procedure is the same for
both:

\begin{enumerate} \def\labelenumi{\arabic{enumi}.}
\tightlist
\item assemble the structures and label each structure (dead or active; active or inactive);
\item compute each item for each structure: the structure-space metrics on the deposited
coordinates, the prediction-based metrics on a Chai-1 \cite{chai2024} prediction of the structure
together with its ligand;
\item compute the AUROC of each item and a bootstrap interval that resamples the clustering unit;
\item rank the items by the direction-free AUROC and compare the top of the ranking with the
best-of-\(M\) permutation threshold.
\end{enumerate}

\subsubsection{2.2.1 Distinguishing natural active and dead enzymes}

\textbf{Dataset.} Pairs of structures of the same protein: the zymogen (dead) and the mature enzyme
(active). The structures were downloaded from the RCSB Protein Data Bank \cite{burley2025,berman2000};
the proteins and the propeptide boundaries come from UniProt \cite{uniprot2023}. The pairs were
assembled in the steps of Table 2, and the evaluation set is the last one:

\begin{longtable}[]{@{}
  >{\raggedright\arraybackslash}p{(\columnwidth - 4\tabcolsep) * \real{0.2143}}
  >{\raggedright\arraybackslash}p{(\columnwidth - 4\tabcolsep) * \real{0.3214}}
  >{\raggedright\arraybackslash}p{(\columnwidth - 4\tabcolsep) * \real{0.4643}}@{}}
\caption*{\textbf{Table 2. Assembly of the zymogen--mature pairs.} From the UniProt search to the
evaluation set of 21 pairs of Results 3.1.1; the last row is the evaluation
set.}\tabularnewline
\toprule\noalign{}
\hc{Step} & \hc{Cases} & \hc{Rule} \\
\midrule\noalign{}
\endfirsthead
\toprule\noalign{}
\hc{Step} & \hc{Cases} & \hc{Rule} \\
\midrule\noalign{}
\endhead
\bottomrule\noalign{}
\endlastfoot
UniProt search &
reviewed proteins of EC 3.4.* with a propeptide feature and PDB structures &
a protein goes on the shortlist when its PDB chain ranges yield both a candidate zymogen and a
candidate mature X-ray entry \\
Proteins on the shortlist & 100 & the proteins that satisfy the rule above \\
Classified PDB entities & 5,717 (1,182 zymogen, 4,528 mature, 7 undetermined) &
each entity is called zymogen or mature on the basis of its SIFTS \cite{dana2019} alignment with
the UniProt chain and the propeptide boundaries, never on the basis of the title of the entry \\
Candidate pairs & 55 pairs from 55 proteins & one zymogen and one mature entity per protein; both
X-ray structures, no engineered mutation, resolution at most 2.5 Å \\
Scored pairs (wider set) & 49 &
structure-space metrics scored on both forms (S3) \\
Re-predicted pairs &
28 & both forms predicted with Chai-1, each with three seeds \\
\textbf{Evaluation set} & \textbf{21} &
the two forms share a ligand; the 7 apo pairs are excluded \\
\end{longtable}

The 21 pairs are 21 proteins, all peptidases (9 serine, 8 cysteine and 4 aspartic; EC 3.4.21,
3.4.22, 3.4.23), with shared ligands such as inhibitors and metal ions. In the wider set of 49
pairs, the median RMSD of the catalytic constellation between the two forms is 0.207 Å (0.484 Å
after backbone superposition). The definitions of the six pair sets used in 3.1 (55, 49, 45, 28, 21
and 96 pairs) and the table that uses each of them are given in Table S4.

\textbf{Why this dataset.} A zymogen and the corresponding mature enzyme keep the catalytic
residues constant, while the enzyme is dead in one form and active in the other; they thus form a
\emph{true} negative: the labels are correct by definition. The subset of 21 pairs is the
evaluation set because it is the only one on which every metric of the main set, structure-space as
well as prediction-based, has a value for both forms of each pair, so that a single n applies to
every row of Table 5. The pairs were selected on the basis of shared ligands and not on the basis
of any metric value. The analysis of 49 pairs is in the Supplement (S3).

\textbf{Scored metrics.} The 33 metrics of the main set (Table 1) are computed on both forms of each
pair. The 29 structure-space metrics are computed on the deposited coordinates over the catalytic
residues, taken from the annotation of M-CSA \cite{ribeiro2018} and UniProt \cite{uniprot2023}. For
the 4 prediction-based metrics, both forms of each pair were re-predicted with Chai-1 together with
the shared ligand (three seeds, averaged), and the active-site accuracy is measured relative to the
own deposited structure of each form. Table S5 lists every metric scored on the 21 pairs, including
the PLACER \cite{anishchenko2025} metrics.

\textbf{Ranking criteria.} Let \(P\) be the dead forms and \(Q\) the active forms, \(|P| = |Q| = n =
21\). The AUROC of item \(m\) is the probability that a randomly chosen dead form has a larger value
than a randomly chosen active form, with ties counting for half \cite{hanley1982}:

\begin{equation*}A_m \;=\; \frac{1}{|P|\,|Q|}\sum_{p \in P}\sum_{q \in Q}\Big[\,\mathbf{1}\big(x_p(m) > x_q(m)\big) + \tfrac{1}{2}\,\mathbf{1}\big(x_p(m) = x_q(m)\big)\Big] \tag{1}\end{equation*}

This equals the Mann--Whitney statistic \(U\) divided by \(|P|\,|Q|\) \cite{mann1947}. \(A_m = 0.5\)
means that the item does not separate the two forms, \(A_m = 1\) that every dead form lies above
every active form and \(A_m = 0\) that every dead form lies below every active form. We ask whether
an item separates the forms, not in which direction, so items are ranked by

\begin{equation*}A^{*}_m \;=\; \max\!\big(A_m,\; 1 - A_m\big) \;\ge\; 0.5 , \tag{2}\end{equation*}

and the directed \(A_m\) is kept so that the direction remains visible (``higher in'' in Table 5 is
the form in which the item takes larger values: dead if \(A_m > 0.5\)). Because \(A^{*}_m\) cannot
fall below 0.5, the lower limit of the interval lies close to 0.5 for almost every item and is not
informative in itself.

The interval is a percentile bootstrap that resamples \textbf{pairs} with replacement, so that both
forms of a drawn pair are included together \cite{efron1979,field2007}. With draws \(b =
1,\dots,N_{\mathrm{boot}}\) (5,000), \(A^{*(b)}_m\) is recomputed on each resampled set and

\begin{equation*}\Big[\,q_{0.025}\big(A^{*(1)}_m,\dots,A^{*(N_{\mathrm{boot}})}_m\big),\;\; q_{0.975}\big(A^{*(1)}_m,\dots,A^{*(N_{\mathrm{boot}})}_m\big)\Big] \tag{3}\end{equation*}

is the 95\% interval. A pair is also a protein, so the N at design level and the N at cluster level
are both 21.

Table 5 ranks the 33 metrics from \(M = 39\) quantities analysed, and the top of such a ranking
looks good by chance. To say how good, the labels are permuted. Within a pair, the two forms are
interchangeable if the item is unrelated to the label, so in replicate \(r = 1,\dots,N_{\mathrm{perm}}\)
we draw \(s_i^{(r)} \sim \mathrm{Bernoulli}(1/2)\) for each pair \(i\), swap the dead and active
labels of the pair when \(s_i^{(r)} = 1\), and recompute \(A^{*(r)}_m\) for each item. The same swaps
are used for all items, which preserves the correlation between metrics. The largest value over the
items in each replicate,

\begin{equation*}T^{(r)} \;=\; \max_{m = 1,\dots,M} A^{*(r)}_{m}, \tag{4}\end{equation*}

has the distribution of the best AUROC of \(M\) unrelated quantities. Its median and 95th
percentile \(q_{0.95}(T)\) form the \textbf{best-of-39 threshold} (median 0.63, 95th percentile
0.68): an observed \(A^{*}_m\) above it is higher than the best of 39 unrelated quantities in 95\%
of the simulations. Taking the maximum over the items is the max-statistic form of permutation
control for multiple comparisons \cite{westfall1993}. The threshold is a reference level for
reading the ranking. For the full table of the metrics (S5), the same procedure over 158 quantities
analysed gives the best-of-158 threshold. In Table 5, AUROC values above the threshold are in bold.
No cut-off is set for this set: we report whether a metric of the main set exceeds the best-of-39
threshold.

\subsubsection{2.2.2 Distinguishing active and inactive de novo designed enzymes}

\textbf{Dataset.} A deposited design campaign for metallohydrolases \cite{kim2026}: 192 designs on
two 96-well plates (two campaigns of 96 designs, the two zinc campaigns with 96 designs that were
also reported with RFdiffusion2 \cite{ahern2026}). Each design had been selected for testing by the
computational filters of its campaign (predicted active-site geometry and preorganisation), and 171
of the 192 were expressed and soluble \cite{kim2026}. Of the 192, \textbf{16 have a reported
kcat/KM} (\emph{active}) and \textbf{176 are inactive here} (no kcat/KM has been reported for them,
and they have not been confirmed as inactive): the 16 are the screening hits that were
characterised with Michaelis--Menten kinetics on purified protein (five from the first campaign,
eleven from the second), and the remaining 176 include the 21 that were not expressed as soluble
protein. The plate screen did not standardise the enzyme concentration and reports no kcat/KM
\cite{kim2026}. The designs lie in \textbf{136 backbone clusters}, and the 16 active designs lie in
8 of them (5 in each of two clusters and one in each of six other clusters). The catalytic motif
consists of the three histidines that coordinate the zinc, derived geometrically and checked
against 23 curated annotations (23 of 23 agree). For the predictions, the ligand is the
transition-state analogue of the design together with the zinc. One set serves for every row of
Table 6: each item is scored on all 192 designs.

\textbf{Why this dataset.} These are the deposited de novo enzyme designs for which both models and
measured activity are available, so that the metrics can be tested on designs and not only on
natural enzymes. A design that is absent from the table of characterised hits has not been reported
as active, and the comparison is active against inactive; this class includes both designs that
were not expressed and designs that were tested without being identified as a hit, and its members
have not been confirmed as inactive. Because every design had already passed the filters of its
campaign, the comparison is made between designs that had been selected for testing. The rank
correlation among the 16 active designs and the hits per plate are in 2.4.2.

\textbf{Scored metrics.} The 31 items are the 29 structure-space metrics of the main set and the two
prediction-based metrics of the main set that are defined on the plates (ipTM and the per-chain pTM
minimum). An item must have a value for at least 80\% of the designs. The AME RMSD and the ligand
clearance are measured on the plates relative to the own model of each design, which is a different
metric from AME relative to a deposited structure (the two variants are never merged), so they are
in the Supplement and not among the ranked items. Table S6 lists every metric scored on the designs.

\textbf{Ranking criteria.} The statistic is Eqs. (1) and (2) with \(P\) the 16 active designs and
\(Q\) the 176 inactive designs, over the 192 designs; ``higher in'' in Table 6 is the group in
which the item takes larger values. The interval resamples the \(G = 136\) \textbf{backbone
clusters} with replacement, with all designs of a drawn cluster included together, so that designs
built on one backbone are not counted as independent (5,000 draws; a draw that contains none of the
16 active designs has no AUROC and is omitted):

\begin{equation*}\Big[\,q_{0.025}\big(A^{*(b)}_m\big),\;\; q_{0.975}\big(A^{*(b)}_m\big)\Big], \qquad A^{*(b)}_m = \max\!\big(A^{(b)}_m,\, 1 - A^{(b)}_m\big), \tag{5}\end{equation*}

with \(A^{(b)}_m\) from Eq. (1) on the designs of the drawn clusters. The N at design level is 192
and the N at cluster level is 136, with 16 active designs in 8 clusters. The chance reference is the
permutation of Eq. (4), in which the activity labels are permuted over the 192 designs instead of
being swapped within pairs, with \(M = 36\) quantities analysed and 5,000 permutations: the
\textbf{best-of-36 threshold} (median 0.66, 95th percentile 0.73), and the best-of-128 threshold
(0.75) over the 128 quantities analysed for Table S6. Because 10 of the 16 active designs lie in two
backbone clusters, designs are not interchangeable and the threshold underestimates the chance
range. Table 6 marks the AUROC values above the best-of-36 threshold in bold.

\begin{center}\rule{0.5\linewidth}{0.5pt}\end{center}

\subsection{2.3 Substrate discrimination test: do prediction-based metrics know which ligand they
were given?}

\textbf{Experiment.} A structure predictor (Chai-1 \cite{chai2024}) is given a protein together with
its own substrate, and then together with a wrong substrate, and the prediction-based quantities are
ranked by how well they distinguish the prediction with the cognate ligand from a prediction with a
wrong ligand for the same enzyme. For example, Chai-1 predicts a hydrolase together with its own
substrate and again with a substrate of an enzyme from a different enzyme class; if ipTM drops when
the wrong substrate is given, the metric ``knows'' which substrate belongs to the protein, and the
AUROC measures how reliably that is the case. Chai-1 is run on the sequence, the ligand (as a
chemical structure) and, where relevant, the catalytic metal, with default settings and a fixed
seed; each prediction yields five models and their confidence scores. Only the ligand that is
presented changes; the protein sequence is the same in every condition, and four conditions are
built for each enzyme (results in 3.2, Table 7):

\begin{enumerate} \def\labelenumi{\arabic{enumi}.}
\tightlist
\item \textbf{cognate}: the substrate of the enzyme (three seeds, 101, 202 and 303; the cognate
condition has two extra seeds that serve as a noise floor and are not used in the evaluation);
\item \textbf{same class}: another substrate of an enzyme from the same enzyme class;
\item \textbf{different class}: a substrate of an enzyme from a different class;
\item \textbf{decoy}: a ligand taken from an unrelated structure.
\end{enumerate}

The wrong ligands each have three seeds, and the catalytic metal is retained in every condition.
The apo condition (no ligand) is excluded for interface and ligand-dependent terms, because without
a ligand there is no interface over which confidence can exist.

\textbf{Dataset.} Natural enzymes from M-CSA \cite{ribeiro2018} with a bound cognate ligand: 59
enzymes enter the experiment, and the \textbf{evaluation set consists of the 55 natural enzymes} for
which every scored quantity has a defined value in the cognate condition and in all three
wrong-ligand conditions, and in which every ligand presented was present in the prediction. The four
exclusions are enzymes whose cognate or wrong ligand contains a haem-type group that cannot be built
as a chemical structure, so that the predictor omits it and the prediction for that cell is
effectively apo. The experiment is balanced: seeds 101, 202 and 303 in every condition. Twenty-eight
of the 59 enzymes are also in the set of 143 enzymes of 2.5.

\textbf{Why this dataset.} A wrong ligand can only be defined for an enzyme whose substrate is
known, and the M-CSA entries in which the cognate reaction ligand is present (150 of the 434
selected entries) are the natural enzymes for which this holds. Natural enzymes carry the main
analysis because their active-site accuracy is measured relative to the deposited structure; the 30
de novo designs, whose active-site accuracy is measured relative to the own model of the design and
never merged with that of the natural enzymes, are in S7.

\textbf{Scored metrics.} Each prediction is assessed using prediction-based quantities, of which
Table 7 ranks 17. The three summaries of the combined Chai-1 score, a rescaled ipTM, are not listed;
their AUROCs are in S7. The 17 listed quantities are the four metrics of the main set from Table 1
(12 rows, one per summary), the AME pass bit and four bookkeeping counters (models, aligned and
catalytic atoms, symmetry swaps) that serve as negative controls. Each score \(s_k\) is returned for
each of the \(K = 5\) models, and each score is summarised over the models as

\begin{equation*}\bar s = \frac{1}{K}\sum_{k=1}^{K} s_k, \qquad s_{\max} = \max_{k} s_k, \qquad s_{\mathrm{SD}} = \sqrt{\frac{1}{K-1}\sum_{k=1}^{K}\big(s_k - \bar s\big)^{2}}, \tag{6}\end{equation*}

and the RMSD also as the minimum over the models, so that one metric appears in several rows of
Table 7, one per summary. Two quantities require a definition of their own (the other confidence
summaries are those of Chai-1 \cite{chai2024}).

\textbf{AME RMSD} (active-site accuracy; adapted from the criterion of the Atomic Motif Enzyme
benchmark \cite{ahern2026}, which superposes on the backbone of the catalytic residues). A model is
superposed on the reference structure with the Kabsch algorithm, on the basis of the C\(\alpha\)
atoms of all residues that it has in common with the reference (the whole shared backbone, and not
only the catalytic residues, which would let the site move freely within the fold and artificially
lower the RMSD). The RMSD of the catalytic atoms is then

\begin{equation*}\mathrm{RMSD}_k \;=\; \sqrt{\frac{1}{L}\sum_{l=1}^{L}\big\lVert \mathbf{y}_l - \mathbf{x}_{l}^{(k)}\big\rVert^{2}}, \qquad \mathrm{AME\ RMSD} = \min_{k} \mathrm{RMSD}_k, \tag{7}\end{equation*}

where \(l = 1,\dots,L\) runs over the heavy atoms of the catalytic residues that retain their
residue type in the prediction, \(\mathbf{y}_l\) is the atom in the reference,
\(\mathbf{x}^{(k)}_l\) the superposed atom of model \(k\), and symmetry-equivalent atoms (for
example the two oxygen atoms of a carboxylate) are assigned so as to give the smaller RMSD. The
reference is the deposited structure for natural enzymes and the own model of the design for
designs; the two runs are never merged. The variant without symmetry resolution is also listed
(``no symmetry resolution'' in Table 7).

\textbf{Ligand clearance} is the shortest distance from the ligand to the backbone atoms N,
C\(\alpha\) and C of the protein, in the most favourable case over the models:

\begin{equation*}C_k \;=\; \min_{a \in \text{ligand}}\ \min_{b \in \{\mathrm{N},\,\mathrm{C}\alpha,\,\mathrm{C}\}} \big\lVert \mathbf{x}_a^{(k)} - \mathbf{x}_b^{(k)}\big\rVert, \qquad \mathrm{clearance} = \max_{k} C_k , \tag{8}\end{equation*}

so that a larger value means more room for the ligand (the AME criterion is \(C > 1.5\) Å).
Interface confidence (ipTM, per-chain pTM minimum) and ligand clearance require a ligand and are
not defined without one.

\textbf{Ranking criteria.} For each quantity \(m\) and each kind of wrong ligand \(c \in
\{\text{same class},\ \text{different class},\ \text{decoy}\}\), the AUROC of Eq. (1) is computed
with the predictions with the cognate ligand as the positive class and the predictions with a wrong
ligand of kind \(c\) as the negative class, one value per system (the mean over the three seeds),
and made direction-free as in Eq. (2); quantities are ranked on the mean of the three,

\begin{equation*}\bar A_m \;=\; \frac{1}{3}\sum_{c}\, A^{*}_{m,c}. \tag{9}\end{equation*}

Each \(A^{*}_{m,c}\) has the percentile bootstrap interval of Eq. (3), which resamples systems
(5,000 draws); ``higher in'' is the condition in which the quantity takes larger values. The chance
reference is the maximum-statistic permutation of Eq. (4): in each of the 5,000 replicates the four
condition labels (cognate, same class, different class, decoy) are randomly permuted within each
system, the three AUROCs and their mean are recomputed for each quantity, and \(T^{(r)} = \max_{m}
\bar A^{(r)}_m\) over the \(M = 30\) quantities analysed; their median and 95th percentile form the
\textbf{best-of-30 threshold} (median 0.56, 95th percentile 0.61) \cite{westfall1993}. Table 7 marks
the mean AUROCs above it in bold. No cut-off is applied; we report which quantities exceed it.

To check whether an effect of ligand size explains the discrimination, the cells (a cognate ligand
and one wrong ligand of an enzyme) are stratified by the difference in the number of heavy atoms,
\(\Delta h = h_{\text{cognate}} - h_{\text{wrong}}\), and in charge, and the paired win fraction of
the quantity is reported (S8):

\begin{equation*}W \;=\; \frac{1}{n_{\mathrm{cells}}}\sum_{\mathrm{cells}}\Big[\mathbf{1}\big(x_{\mathrm{cognate}} > x_{\mathrm{wrong}}\big) + \tfrac12\,\mathbf{1}\big(x_{\mathrm{cognate}} = x_{\mathrm{wrong}}\big)\Big]. \tag{10}\end{equation*}

\begin{center}\rule{0.5\linewidth}{0.5pt}\end{center}

\subsection{2.4 Activity ranking test: do the metrics track the measured activity?}

\textbf{Experiment.} We ask whether the order of real variants (or of the 16 designs with a measured
activity) according to a metric agrees with their order according to the measured activity. For
example, the variants are lined up by how strongly the assay scored each of them as impaired and
again by how much a metric changes for each of them; if the two orders agree, the metric ranks the
variants as the assay does, and the rank correlation measures how well. This is a claim about
ranking, so the statistic is a rank correlation; the AUROC of the binary contrast (impaired versus
wild-type-like variants) is given for BglB in the Supplement (S9). Two datasets are used, each with
its own table and its own n: 432 BglB variants (2.4.1; results in 3.3.1, Table 8) and the 16 plated
designs with a measured activity (2.4.2; results in 3.3.2, Table 9); the other 176 designs of the
campaign have no measured activity and are not ranked.

\subsubsection{2.4.1 Ranking the activity of natural enzymes}

\textbf{Dataset.} β-Glucosidase B (BglB) from \emph{Paenibacillus polymyxa}, characterised by the
D2DCure consortium \cite{carlin2016} (data retrieved from d2dcure.com on 2026-08-06; the licence is
not stated, so variant-level data are not distributed further). The variant numbering is shifted by
three positions relative to the numbering of the crystal structure (RCSB PDB entry 2JIE
\cite{burley2025,berman2000}; structure position = variant position + 3). The catalytic residues are
Glu167 (proton donor) and Glu356 (nucleophile) in structure numbering, taken from the active-site
features in UniProt \cite{uniprot2023}, the conserved motifs TINEP and ITENG, and the covalent bond
to the glucosyl intermediate in the deposited structure. The funnel from the measured rows to the
evaluation set is in Table 3:

\begin{longtable}[]{@{}
  >{\raggedright\arraybackslash}p{(\columnwidth - 4\tabcolsep) * \real{0.2075}}
  >{\raggedright\arraybackslash}p{(\columnwidth - 4\tabcolsep) * \real{0.3396}}
  >{\raggedright\arraybackslash}p{(\columnwidth - 4\tabcolsep) * \real{0.4528}}@{}}
\caption*{\textbf{Table 3. From the D2DCure rows to the evaluation set of 432 variants.} The funnel
for Results 3.3.1; the last row is the set on which each row of Table 8 is
computed.}\tabularnewline
\toprule\noalign{}
\hc{Step} & \hc{Cases} & \hc{Rule} \\
\midrule\noalign{}
\endfirsthead
\toprule\noalign{}
\hc{Step} & \hc{Cases} & \hc{Rule} \\
\midrule\noalign{}
\endhead
\bottomrule\noalign{}
\endlastfoot
Measured rows &
1,288 rows, of which 272 are wild-type replicates measured at eight institutes & the
D2DCure characterisation data \\
Single-point variants & 655 at 207 positions & the variants from the characterisation data \\
Set aside & 108 destabilised, 83 not expressed, 32 with insufficient data & a
variant is \textbf{folded} if it was expressed and is not destabilised (T50 or
Tm less than 2 standard deviations below the wild type); the kinetics of the others cannot be
attributed to the catalytic site \\
Folded variants with kinetics & 432 at 175 positions: 166 impaired, 254 wild-type-like, 12
improved &
\textbf{impaired} if kcat/KM is at least 2 standard deviations (a 3.8-fold loss or more) below
the wild type, \textbf{wild-type-like} if it is within 2 standard deviations, \textbf{improved} if
it is above \\
\textbf{Evaluation set} & \textbf{432 variants} & one set serves for every row of Table 8; a
structure-space metric has a value for 431 of the 432 variants and a prediction-based metric for
all 432 \\
\end{longtable}

The spread of the wild-type replicates (0.293 log10) forms the noise floor, and each variant is
compared with the wild type of the same institute. The binary contrast of the Supplement (S9) uses
the 420 variants that are impaired or wild-type-like. Substrate: the prediction-based metrics use
the assay substrate 4-nitrophenyl-β-D-glucopyranoside (pNPG), the reporter substrate of the
published BglB assays \cite{carlin2016,carlin2017}, and no metal, because BglB is metal-independent.

\textbf{Why this dataset.} BglB provides hundreds of single-point variants of one enzyme, with
kinetics, expression and thermal stability (T50, Tm) measured within one consortium, so that a claim
about ranking can be tested on a scale that no set of designs offers, and so that a loss of activity
through a folding failure (a destabilised or non-expressed variant) can be distinguished from a loss
in the catalytic site. BglB is a single metal-independent glycosidase and its 432 variants are
single substitutions of that one protein; ranking point mutants of one enzyme is a different task
from ranking de novo designs, which differ from one another in many residues and in the backbone
(2.4.2).

\textbf{Scored metrics.} Each variant is built in silico with the same substitution in rotamer
space, the same repacking and the same minimisation as in 2.5 (the wild type undergoes the same
treatment with five seeds), and the metrics are computed for each variant over the two catalytic
glutamates. The prediction-based metrics are computed on Chai-1 predictions of each variant and of
the wild type with the assay substrate; AME is measured relative to the wild-type crystal, which
carries a covalent glucosyl intermediate and not the substrate. The 32 ranked items of Table 8 are
the 28 structure-space metrics of the main set that BglB defines (the omega term does not change
with a side-chain substitution on a fixed backbone) and the 4 prediction-based metrics of the main
set.

\textbf{Ranking criteria.} The measured \textbf{impairment} of variant \(v\) is the log ratio of its
kcat/KM relative to the wild-type value of the same institute,

\begin{equation*}I_v \;=\; \log_{10}\frac{\kappa_{\mathrm{wt}}}{\kappa_{v}}, \tag{11}\end{equation*}

larger for a more strongly impaired variant (\(\kappa\) is kcat/KM). A metric enters as the size of
its change, the absolute difference between the variant and the treated unsubstituted protein,
\(\lvert\Delta_v(m)\rvert = \lvert m(v) - m(\mathrm{wt})\rvert\). The direction is a larger change
in a more strongly impaired variant, so the directed Spearman correlation is reported and a negative
value is left unchanged. With \(r_v\) the rank of the score of the item and \(s_v\) the rank of
\(I_v\) among the \(n = 432\) variants (average rank for ties), the Spearman rank correlation is the
Pearson correlation of the ranks \cite{spearman1904}:

\begin{equation*}\rho_m \;=\; \frac{\sum_{v}(r_v-\bar r)(s_v-\bar s)}{\sqrt{\sum_{v}(r_v-\bar r)^2\;\sum_{v}(s_v-\bar s)^2}} \tag{12}\end{equation*}

\(\rho = 1\) means that the item orders the variants exactly as the assay does, \(\rho = 0\) that
the two orderings are unrelated, and \(\rho = -1\) that they are opposite. The 95\% interval is a
percentile bootstrap (Eq. 3 with \(\rho\) instead of \(A^{*}\)) that redraws the 175
\textbf{positions} with replacement (5,000 draws), because variants at one position are not
independent; the N at design level is 432 and the N at cluster level is 175. Table 8 ranks the 32
metrics from \(M = 41\) quantities analysed by \(\rho_m\), and the top of such a ranking looks good
by chance, so the measured impairments are permuted over whole positions with the same number of
variants (a block permutation, so that the variants at one position keep their impairments together,
as in the bootstrap), \(\rho^{(r)}_m\) is recomputed for each item with the same permutation, and

\begin{equation*}T^{(r)} \;=\; \max_{m = 1,\dots,M} \rho^{(r)}_{m}, \qquad r = 1,\dots,N_{\mathrm{perm}} . \tag{13}\end{equation*}

The median and 95th percentile of \(T\) form the \textbf{best-of-41 threshold}: the correlation
that the best of 41 unrelated quantities would exceed in 5\% of the permutations (median 0.24, 95th
percentile 0.32; for the full table S9, over all 114 quantities analysed, 0.32). It is a reference
level for reading the ranking; a \(\rho\) above it is in bold in Table 8. The signed correlation
(S10) and the AUROC of the binary contrast (S9) are given alongside. The 15 prediction-based
metrics, with the columns for the signed correlation of the change and the AUROC, are in Table S11.

\subsubsection{2.4.2 Ranking the activity of the 16 de novo designs with a reported kcat/KM}

\textbf{Dataset.} The campaign of 192 designs of 2.2.2 \cite{kim2026}. The ranking is made on 16
designs, not on the 192: only the 16 designs with a reported kcat/KM have a measured activity with
which an item can be correlated, so the \textbf{evaluation set consists of those 16 designs} and one
n serves for every row of Table 9. The other 176 designs have no measured activity and are not
ranked. They are the screening hits of the two campaigns that were characterised with
Michaelis--Menten kinetics (five and eleven), so the evaluation set is a selection of the designs
that showed activity in the screen, and the reported kcat/KM spans 35 to 53,000
M\textsuperscript{−1} s\textsuperscript{−1} \cite{kim2026}. They lie in 8 of the 136 backbone
clusters, so the effective number of independent designs is smaller than 16. The 192 designs are
used for the AUROC of 2.2.2 (Table 6) and for the filter-style analysis (S12), which are analyses of
active against inactive and not rankings of the designs by activity.

\textbf{Why this dataset.} The same designs as in 2.2.2, so that discrimination (active against the
rest of the plate, 192 designs) and ranking (among the 16 active designs, the only designs with a
measured activity) are two readings of one campaign. The question that a designer puts to a metric
is whether it would have enriched the hits on a plate, and this is a deposited design campaign on
which that can be asked. The AUROC is reported in 2.2.2.

\textbf{Scored metrics.} The 31 items of Table 9 are those of Table 6: the 29 structure-space
metrics of the main set and the two prediction-based metrics of the main set that the plates define
(ipTM and the per-chain pTM minimum). Designs are scored as deposited, and their predictions use the
transition-state-analogue ligand and the zinc of the design, with the accuracy measured relative to
the own model of the design. AME and the ligand clearance are in the Supplement (S13) and are not
among the ranked items, as in 2.2.2.

\textbf{Ranking criteria.} The correlation is Eq. (12) between the original value of the item on the
design and \(\log_{10}\mathrm{kcat/KM}\) of the \(n = 16\) designs that have a reported value; no
direction is assumed, so items are ranked on \(\lvert\rho_m\rvert\) and the sign is reported
(``higher value means higher or lower activity''). The interval is a 90\% percentile bootstrap that
redraws designs (2,000 draws).

A \(\rho\) is in bold in Table 9 when the 90\% interval excludes 0. The filter-style analysis does
not rank the designs by activity: it uses the active and inactive labels of all 192 designs and
fills a 96-well plate only with the quarter of the 192 designs with the highest, or with the lowest,
values of the item: with \(n_{\mathrm{keep}} = 48\) designs kept, of which \(k\) are active, the
expected hits per plate are

\begin{equation*}\text{hits per plate} \;=\; 96\,\frac{k}{n_{\mathrm{keep}}} \;=\; 2k , \tag{14}\end{equation*}

against \(96 \times 16/192 = 8.0\) for an unfiltered plate. The band of a random filter is the
distribution of \(k\) when 48 designs are chosen at random, hypergeometric with 16 active designs
among 192,

\begin{equation*}\Pr(k > h) \;=\; \sum_{j > h}\frac{\binom{16}{j}\binom{176}{48-j}}{\binom{192}{48}} , \tag{15}\end{equation*}

so that 95\% of random choices contain at most 14.0 hits per plate (\(h = 7\) active designs, the
smallest \(h\) with \(\Pr(k > h) \le 0.05\)), a random choice contains more with probability
\(\Pr(k > 7) = 0.022\) and either end of a random item does so with probability 0.044 (S12). The
correlation of each metric is in Table S13, and the enrichment at keep fractions of 5, 10, 25 and
50\% in both directions is in Table S14 (structure-space metrics) and Table S15 (prediction-based
metrics).

\begin{center}\rule{0.5\linewidth}{0.5pt}\end{center}

\subsection{2.5 Detection ability test: do the metrics respond more strongly to a catalytic lesion
than to a matched control?}

\textbf{Experiment.} We damage the catalytic residues of natural enzymes, in the structure for the
structure-space metrics and in the sequence of a re-predicted complex for the prediction-based
metrics, and ask whether each metric changes more strongly than when the same damage is applied at a
matched non-catalytic site of the same protein. We call each such deliberate change a
\emph{lesion}. In a serine protease, for example, the catalytic triad is Asp, His and Ser. The
\emph{Ala step} builds a structure in which those three residues are replaced by Ala. The
\emph{control} builds a structure in which three buried, non-catalytic residues of the same original
types are replaced by Ala, with the same repacking. A metric is \emph{specific} if it changes
clearly more in the first structure than in the second. Two kinds of lesion are applied, each with
its own datasets: the catalytic lesion (the chemistry ladder, 2.5.1; results in 3.4.1, Table 10) and
the second-shell lesion (2.5.2; results in 3.4.2).

\textbf{Change.} Each change is applied in the side-chain rotamer space on the fixed backbone,
followed by a fixed-backbone repacking of 8 Å, with the changed residues held fixed, and a
restrained minimisation (PyRosetta \cite{chaudhury2010}); the protonation is fixed. The
\textbf{chemistry ladder} replaces each catalytic residue in one of four steps of increasing change
in side-chain volume: \textbf{isosteric} (for example Ser→Cys, Asp→Asn, Thr→Val; median change
15.3 Å³), \textbf{non-isosteric} (36.4 Å³), \textbf{to Ala} (53.2 Å³) and \textbf{to Gly}
(80.5 Å³). \textbf{Second-shell removal} mutates 1, 2 or 4 residues in the shell around the
catalytic residues. The \textbf{detection floor} replaces all catalytic residues at once by Gly,
which is the same lesion as the Gly step, and asks whether the metric notices this. Table S16 gives
the result of the detection floor and of the other conditions of the sanity floor by class of
metric, and checks that the detection floor and the Gly step are identical. For the
prediction-based metrics, the same four steps of the chemistry ladder are applied in the
\textbf{sequence}: for each prediction one catalytic residue is substituted and the complex is
re-predicted with Chai-1 \cite{chai2024} (one seed, 101).

\textbf{Comparison.} Two comparisons are made, and they answer different questions.

\begin{itemize}
\tightlist
\item \emph{Against the structure's own reference.} Each perturbed structure and its reference
undergo an identical treatment, because some metrics change through the treatment alone; each arm is
compared with its own treated reference.
\item \emph{Against the matched control.} The control makes the \textbf{identical substitution}
(same original and same new residue type) at buried, non-catalytic positions of the same protein,
matched on degree of burial (within 0.10 relative solvent-accessible surface area) and, in the
primary analysis, on a local packing band. When no exact match exists, the matcher relaxes its
criteria in a fixed order and records the loosest level used (the loosest level needed for each
lesion test is in Table S17). Only comparisons in which both arms change the same number of residues
are counted (the balance of the numbers of changed residues in the two arms is in Table S18). For the
re-predicted ladder, each substitution is paired with the identical substitution at a buried
non-catalytic position (same original and same new residue, matched on degree of burial), and the
change is measured relative to the unperturbed prediction of the same system at the same seed.
\end{itemize}

\textbf{Response.} For an item \(m\), an enzyme \(i\) and an arm \(a \in \{\mathrm{cat},
\mathrm{ctl}\}\) (the catalytic arm and the control arm), the response at step \(s\) is the change
of the item relative to its own treated reference,

\begin{equation*}\Delta^{a}_{i}(m, s) \;=\; m\big(\text{changed}^{a}_{i,s}\big) - m\big(\text{reference}^{a}_{i}\big) , \tag{16}\end{equation*}

summarised as the median over enzymes. For the structure-space metrics, the site-bound items are
computed over the changed residues in each arm (the catalytic residues in the catalytic arm, the
control residues in the control arm); in second-shell removal both arms are scored over the
catalytic residues. For the prediction-based metrics, both members of the pair are scored over the
catalytic set \emph{minus} the substituted position, so that the active-site accuracy reflects only
the displacement of the remaining catalytic residues. An item \textbf{responds} at a step when the
95\% interval of its median response excludes zero.

\subsubsection{2.5.1 Detection of a catalytic lesion}

\textbf{Dataset.} The datasets of this test are listed in Table 4:

\begin{longtable}[]{@{}
  >{\raggedright\arraybackslash}p{(\columnwidth - 4\tabcolsep) * \real{0.2407}}
  >{\raggedright\arraybackslash}p{(\columnwidth - 4\tabcolsep) * \real{0.3333}}
  >{\raggedright\arraybackslash}p{(\columnwidth - 4\tabcolsep) * \real{0.4259}}@{}}
\caption*{\textbf{Table 4. Datasets of the catalytic lesion test.} The sets on which the catalytic
lesion is scored in Results 3.4.1 and the sets that are excluded or used for sensitivity
analyses.}\tabularnewline
\toprule\noalign{}
\hc{Dataset} & \hc{Cases} & \hc{How the set is built and what it is used for} \\
\midrule\noalign{}
\endfirsthead
\toprule\noalign{}
\hc{Dataset} & \hc{Cases} & \hc{How the set is built and what it is used for} \\
\midrule\noalign{}
\endhead
\bottomrule\noalign{}
\endlastfoot
M-CSA source \cite{ribeiro2018} &
434 entries of the M-CSA snapshot of 2026-07-31 selected for this study & the curated mechanisms
and catalytic residues of the Mechanism and Catalytic Site Atlas; structures come from the Protein
Data Bank \cite{burley2025,berman2000} \\
143-enzyme set (primary) &
\textbf{143 enzymes}, 77 enzyme sub-subclasses, median 6 catalytic residues & at most 25 per
top-level EC class (25, 25, 25, 25, 25 and 18 enzymes), at least 2 resolved catalytic residues,
ranked within a class by quality, number of catalytic residues and resolution; packing-matched
controls with count gating \\
53-enzyme set & 53 enzymes (35 peptidases, 34 with a ligand), of which 52 are not in the 143-enzyme
set &
active sites from UniProt \cite{uniprot2023} and M-CSA; one enzyme is the same protein as an enzyme
of the main set (PDB entries 1JF8 and 1LJL) and is counted once; controls matched only on residue
type and degree of burial \\
\textbf{Evaluation set for structure space} &
\textbf{195 enzymes} (143 + 52) & the 29 structure-space metrics of the main set; the main set
alone is kept alongside the pooled score (Table S19) \\
\textbf{Evaluation set for prediction-based metrics} &
\textbf{59 natural enzymes, 286 catalytic/control pairs} & the natural arm of the re-predicted
ladder (the 59 natural enzymes of 2.3, of which 28 belong to the 143) \\
Excluded from pooling & 30 de novo designs; 15 designed systems of the ladder (74 systems,
312 pairs with the 59 natural ones) &
the control of a design is constructed, so designs are not pooled (the designs form
the de novo sensitivity set of Table S20) \\
Shortened-motif set & the 143 enzymes with at most 3 retained catalytic residues & a
sensitivity analysis of the specificity ratio (S21) \\
\end{longtable}

\textbf{Why this dataset.} A lesion at a catalytic residue only makes sense for enzymes whose
catalytic residues have been curated, and M-CSA is the curated source. The chemistry ladder isolates
chemical identity at almost constant geometry, and the isosteric step is the hardest synthetic
negative case; the matched control, which is mandatory at every level, tests whether a metric knows
that the lesion is at the catalytic site and not merely that a buried residue has changed. The 52
enzymes of the 53-enzyme set that are not in the 143-enzyme set are pooled with it; adding them
changes the mean R of a metric by at most 0.06 (median 0.02, Table S19). The prediction-based arm
uses the natural enzymes of the substrate-swap set, which have a cognate ligand, because the
interface metrics are not defined without a ligand.

\textbf{Scored metrics.} The 29 structure-space metrics (Table 1) are computed on each structure.
The four prediction-based metrics of the main set (Table 1, defined in 2.3) are computed on each
re-predicted complex. The seed noise of the ladder is estimated from a replicate with a second seed
(Table S22). The 33 ranked items of Table 10 are the 29 structure-space metrics and the 4
prediction-based metrics of the main set; the 8 other rows (6 supplementary metrics, which are
per-residue variants and aggregation variants, and 2 aliases of ranked metrics) and the per-step
flags are in Table S19.

\textbf{Ranking criteria.} The \textbf{rank statistic R} is the probability form of the comparison
between the two arms. It is the quantity on which a shortlist depends, because pipelines rank
candidates and do not use absolute magnitudes; Results 3.4 uses it to rank the metrics by lesion.
For an item \(m\), a step \(s\) and the \(n_s\) enzymes that contain both arms and in which at least
one arm changes,

\begin{equation*}R_{m,s} \;=\; \frac{1}{n_s}\sum_{i=1}^{n_s}\Big[\mathbf{1}\big(\lvert\Delta^{\mathrm{cat}}_{i}\rvert > \lvert\Delta^{\mathrm{ctl}}_{i}\rvert\big) + \tfrac{1}{2}\,\mathbf{1}\big(\lvert\Delta^{\mathrm{cat}}_{i}\rvert = \lvert\Delta^{\mathrm{ctl}}_{i}\rvert\big)\Big] \tag{17}\end{equation*}

(an enzyme in which both changes are exactly zero contains no ranking information and is omitted; a
step requires at least five enzymes). \(R = 0.5\) means no specificity, \(R > 0.5\) that the item
changes more at the catalytic lesion, \(R < 0.5\) that it changes more at the control; unlike a
direction-free AUROC, it has a lower limit of 0 and not of 0.5.

\emph{Which steps count.} When the control arm does not change beyond the item's own noise, \(R\)
measures whether the item changes at all and not whether it is specific, and when the catalytic
response cannot be distinguished from zero, there is nothing to rank. A structure-space step is
therefore only included in the set \(S_m\) of counted steps if the item responds at that step (the
interval of its median change excludes zero) and its control response lies above its noise, as
established on the 143 main enzymes. The prediction-based cells have no such flag, because the
control response of the predictor is limited by noise; they all count. The score of an item is the
mean over the counted steps,

\begin{equation*}\bar R_m \;=\; \frac{1}{\lvert S_m\rvert}\sum_{s \in S_m} R_{m,s} , \tag{18}\end{equation*}

and an item without a counted step has no mean. The counted steps differ between items, so the
means of different items are taken over different sets of steps and are not strictly comparable; the
number of counted steps for each item, \(\lvert S_m\rvert\), is given next to its mean in Table 10
and, set against the steps with a value, in Table S19.

\emph{Interval.} A 95\% percentile bootstrap interval of \(\bar R_m\) (Eq. 3 with \(\bar R_m\)
instead of \(A^{*}\); 5,000 draws) that resamples enzyme sub-subclasses, the first three EC fields,
a stated proxy for superfamily and, for the predictor, the systems; the same drawn clusters are used
for every step of an item.

\emph{Chance reference.} As in 2.2, the top of a ranking of \(M\) quantities looks good by chance.
The catalytic and control labels are randomly swapped within each enzyme (\(s_i^{(r)} \sim
\mathrm{Bernoulli}(1/2)\), \(N_{\mathrm{perm}} = 5{,}000\) replicates, the same swaps for every
item), \(\bar R^{(r)}_m\) is recomputed for each ranked item, and \(T^{(r)} = \max_m \bar
R^{(r)}_m\) (Eq. 4) has the distribution of the best score of \(M\) unrelated quantities; its median
and 95th percentile form the \textbf{best-of-33 threshold} (\(M = 33\) quantities analysed; median
0.57, 95th percentile 0.62), a reference level for reading the ranking \cite{westfall1993}. A mean
above it is in bold in Table 10.

\emph{Onsets and grouping.} The metrics are also distinguished by the lesion that they can detect.
The steps are ordered isosteric \textless{} non-isosteric \textless{} Ala \textless{} Gly (15.3,
36.4, 53.2 and 80.5 Å³). The \textbf{onset of detection} is the smallest step at which the item
responds (Eq. 16; for the structure-space metrics the response analysis of the 143 main enzymes,
2,000 draws; for the prediction-based metrics the signed changes of the natural pairs, bootstrap over
systems),

\begin{equation*}s_{\mathrm{det}}(m) \;=\; \min\{\,s : m \text{ responds at } s\,\}, \tag{19}\end{equation*}

with ``Never'' if there is none. Rows of Table 10 are grouped by \(s_{\mathrm{det}}\) (the first
column), ranked within each group by \(\bar R_m\), and a metric without a counted step closes its
group. The \textbf{onset of specificity} is the smallest step at which the item is specific
according to the rule for the specificity ratio (structure-space metrics only; listed in Table S19);
the ratio and the rule are

\begin{equation*}\mathrm{SR}_{m,s} \;=\; \frac{\operatorname{median}_i \lvert\Delta^{\mathrm{cat}}_{i}\rvert}{\operatorname{median}_i \lvert\Delta^{\mathrm{ctl}}_{i}\rvert}, \tag{20}\end{equation*}

and an item is called \textbf{specific} only when all five conditions are met: (1) the catalytic
response is significant; (2) the interval of SR excludes 1; (3) the control arm provides at least
half the number of observations of the catalytic arm; (4) the control response exceeds 1\% of the
item's own native standard deviation, because a ratio whose denominator lies below the item's noise
is not specificity; (5) both arms changed an equal number of residues (2,000 draws; counts are given
as distinct metrics, with literal duplicates merged). The specificity ratio, its interval and the
verdict (specific, non-specific, blind or invariant) of each eligible metric at each step are in
Table S23, the cells that pass the controls are listed in Table S24; and the shape of R per step on
the 143 main enzymes only, with 90\% intervals, is in Table S25.

\emph{Specificity ratio of the predictor.} For the prediction-based metrics, the ratio of the mean
change at the catalytic residue to the mean change at the control,
\(\overline{\lvert\Delta\rvert}_{\mathrm{cat}} / \overline{\lvert\Delta\rvert}_{\mathrm{ctl}}\) (a
median-based form is also given), is computed at each step with a bootstrap over systems (5,000
draws); the active-site accuracy is reported as a secondary displacement without a specificity
verdict, because the substituted residue is excluded from its own comparison. The panel of the main
set is summarised in the Supplement by the geometric mean of SR over the metrics with a defined
ratio, with a nested bootstrap (3,000 draws) that recomputes each metric within each draw, so that
the correlation between metrics is taken into account; equivalence to 1 is judged against a band of
±25\% (S26). The re-predicted ladder at each step, for all metrics, natural and de novo, is in Table
S27.

\subsubsection{2.5.2 Detection of the second-shell lesion}

\textbf{Dataset.} The 143 main enzymes of 2.5.1 pooled with the 52 enzymes of the 53-enzyme set that
are not in the 143-enzyme set, with a control arm for at most 193 of them; structure-space metrics
are scored on this lesion. The number of enzymes in which an item changes at all, and thus carries
information, ranges from 0 to 182 per item and dose.

\textbf{Why this dataset.} The second-shell lesion asks whether the metrics look beyond the first
shell: second-shell residues around the catalytic residues are removed in graded doses of 1, 2 and 4
residues, and the control removes the same number of residues around a matched non-catalytic site.
The same enzymes as in the catalytic lesion are used, so that the two lesions can be read on one
panel and in one population, and the same pooling applies.

\textbf{Scored metrics.} The 29 structure-space metrics of the main set, scored over the catalytic
residues in both arms (so that an item with site scope never knows which residues are catalytic).
The 8 other rows (6 supplementary metrics and 2 aliases of ranked metrics) and the per-dose flags
are in Table S28.

\textbf{Ranking criteria.} The doses are \(d \in \{1, 2, 4\}\) removed residues instead of the steps
of Eq. (17), and the rank statistic is

\begin{equation*}R_{m,d} \;=\; \frac{1}{n_d}\sum_{i=1}^{n_d}\Big[\mathbf{1}\big(\lvert\Delta^{\mathrm{cat}}_{i}\rvert > \lvert\Delta^{\mathrm{ctl}}_{i}\rvert\big) + \tfrac{1}{2}\,\mathbf{1}\big(\lvert\Delta^{\mathrm{cat}}_{i}\rvert = \lvert\Delta^{\mathrm{ctl}}_{i}\rvert\big)\Big], \qquad \bar R_m = \frac{1}{3}\sum_{d} R_{m,d} . \tag{21}\end{equation*}

with the same bootstrap over enzyme sub-subclasses, the same onset of detection (Eq. 19, over doses)
and the same grouping as for the catalytic lesion. The doses with a response and with a control
response above the item's noise are counted for each item, and a dose only enters a counted set when
both hold, as in 2.5.1; here no cell qualifies for any item, so the ranking uses the raw
\(R_{m,d}\) of each dose with at least five informative enzymes, raw values are reported and no
best-of-\(M\) threshold is used. Results 3.4.2 reports the raw \(\bar R_m\) of the items and the
number of doses (of 3) at which each item responds and at which its control response lies above its
noise; the values at each dose are in Table S28.

\section{3 Results}

Each subsection follows the same order: the experiment, the dataset, the criteria by which the
metrics are ranked and the results table. Numbers of metrics are numbers of distinct metrics in the
main set of 33 (29 structure-space metrics, 4 prediction-based metrics; Table 1 of the Methods), and
intervals are 95\% bootstrap intervals that resample the unit named in each subsection, unless
stated otherwise. Numbers in square brackets are references, listed in the References. Full
per-metric tables and further analyses are in the supplementary tables (S-numbers), which are
described in the Supplement PDF and released as CSV files (\texttt{main/} and
\texttt{supplementary/}).

\begin{center}\rule{0.5\linewidth}{0.5pt}\end{center}

\subsection{3.1 Activity discrimination test}

\textbf{Experiment.} Metrics are scored on real proteins in which one state is catalytically dead
and the other active, and each metric is ranked by how well its value separates the two states
(Methods 2.2). Two datasets are used, each with its own table.

\subsubsection{3.1.1 Distinguishing natural active and dead enzymes}

\textbf{Dataset.} 21 zymogen--mature enzyme pairs from 21 peptidases (9 serine, 8 cysteine and 4
aspartyl peptidases), downloaded from the RCSB Protein Data Bank \cite{burley2025,berman2000}. The
zymogen is the dead form and the mature enzyme the active form, and both forms share a ligand.
Catalytic residues come from M-CSA \cite{ribeiro2018} and UniProt \cite{uniprot2023} annotation;
prediction-based quantities are Chai-1 \cite{chai2024} predictions of both forms with the shared
ligand. Each item is scored on all 21 pairs (Methods 2.2.1).

\textbf{Ranking criteria.} The 33 metrics of the main set are ranked on the direction-free AUROC
\cite{hanley1982,mann1947} of dead against active (0.5 = no separation), with a 95\% bootstrap
interval that resamples pairs. ``Higher in'' is the form in which the metric takes larger values. A
bold AUROC would exceed the best-of-39 threshold \cite{westfall1993}, the 95th percentile of the
best AUROC among 39 unrelated quantities under random label swaps within pairs (median 0.63, 95th
percentile 0.68).

\textbf{Result.} None of the 33 metrics exceeds the best-of-39 threshold. The highest AUROC is 0.61,
for the intra-residue repulsion of the catalytic residues. The AUROCs of the 26 metrics shown lie
between 0.52 and 0.61, and their 95\% intervals all contain the range from 0.53 to 0.58.

\begin{longtable}[]{@{}
  >{\raggedright\arraybackslash}p{(\columnwidth - 10\tabcolsep) * \real{0.0709}}
  >{\raggedright\arraybackslash}p{(\columnwidth - 10\tabcolsep) * \real{0.2992}}
  >{\raggedright\arraybackslash}p{(\columnwidth - 10\tabcolsep) * \real{0.1890}}
  >{\raggedright\arraybackslash}p{(\columnwidth - 10\tabcolsep) * \real{0.1260}}
  >{\raggedright\arraybackslash}p{(\columnwidth - 10\tabcolsep) * \real{0.2047}}
  >{\raggedright\arraybackslash}p{(\columnwidth - 10\tabcolsep) * \real{0.1102}}@{}}
\caption*{\textbf{Table 5. Separation of real dead and active enzymes on the evaluation set of 21
pairs (the 26 highest-ranked of 33 metrics).} The 33 metrics of the main set are ranked on the
direction-free AUROC of dead against active (lower limit 0.5) over 21 pairs from 21 proteins; each
row has n = 21 pairs. Intervals are 95\% bootstrap intervals that resample pairs. ``Higher in'' is
the form in which the metric takes larger values. A bold AUROC would exceed the best-of-39 threshold,
the 95th percentile of the best AUROC among 39 unrelated quantities in label-swap simulations
(median 0.63, 95th percentile 0.68; defined in the Methods; a reference level for reading the
ranking); none does. The seven lowest-ranked metrics (AUROC 0.50 to 0.51) and every metric scored on
these pairs are in Table S5. Source: main/T5\_dead\_vs\_active.csv.}\tabularnewline
\toprule\noalign{}
\hc{Rank} & \hc{Item} & \hc{Group} & \hc{Mode} & \hc{AUROC {[}95\% interval{]}} & \hc{Higher in} \\
\midrule\noalign{}
\endfirsthead
\toprule\noalign{}
\hc{Rank} & \hc{Item} & \hc{Group} & \hc{Mode} & \hc{AUROC {[}95\% interval{]}} & \hc{Higher in} \\
\midrule\noalign{}
\endhead
\bottomrule\noalign{}
\endlastfoot
1 & Intra-residue repulsion (Rosetta, site) \cite{alford2017} & Rosetta energy &
structure & 0.61 {[}0.52, 0.72{]} & dead \\
2 & Repulsion (Rosetta, site) \cite{alford2017} & Rosetta energy &
structure & 0.60 {[}0.51, 0.72{]} & dead \\
3 & Fraction buried & Geometry &
structure & 0.57 {[}0.50, 0.69{]} & dead \\
4 & Attraction (Rosetta, site) \cite{alford2017} & Rosetta energy &
structure & 0.56 {[}0.51, 0.64{]} & active \\
5 & lk-ball solvation (Rosetta, site) \cite{alford2017} & Rosetta energy &
structure & 0.56 {[}0.50, 0.66{]} & dead \\
6 & Ramachandran (Rosetta, site) \cite{alford2017} & Rosetta energy &
structure & 0.56 {[}0.51, 0.63{]} & dead \\
7 & Relative SASA, mean & Geometry &
structure & 0.55 {[}0.50, 0.66{]} & active \\
8 & Total SASA & Geometry & structure & 0.55 {[}0.50, 0.67{]} & active \\
9 & Rotamer, Dunbrack (Rosetta, site) \cite{alford2017} & Rosetta energy &
structure & 0.55 {[}0.50, 0.66{]} & dead \\
10 & Pairwise distance, max & Geometry & structure & 0.54 {[}0.50, 0.64{]} & dead \\
11 & ipTM (Chai-1) \cite{chai2024,evans2021} & Interface confidence &
prediction & 0.54 {[}0.50, 0.61{]} & active \\
12 & pKa, range (PROPKA) \cite{olsson2011,sondergaard2011} & pKa & structure &
0.54 {[}0.50, 0.69{]} & dead \\
13 & Polar contacts & Geometry &
structure & 0.54 {[}0.50, 0.59{]} & dead \\
14 & Ligand clearance \cite{ahern2026} & Active-site accuracy & prediction &
0.54 {[}0.50, 0.65{]} & dead \\
15 & Net van der Waals (Rosetta, site) \cite{alford2017} & Rosetta energy &
structure & 0.54 {[}0.50, 0.61{]} & active \\
16 & Relative SASA, max & Geometry &
structure & 0.53 {[}0.50, 0.65{]} & active \\
17 & Radius of gyration & Geometry & structure & 0.53 {[}0.50, 0.64{]} & dead \\
18 & AME RMSD \cite{ahern2026} & Active-site accuracy & prediction &
0.53 {[}0.50, 0.66{]} & active \\
19 & Pairwise distance, mean & Geometry & structure & 0.53 {[}0.50, 0.64{]} & dead \\
20 & Solvation (Rosetta, site) \cite{alford2017} & Rosetta energy &
structure & 0.53 {[}0.50, 0.60{]} & dead \\
21 & Omega torsion (Rosetta, site) \cite{alford2017} & Rosetta energy &
structure & 0.53 {[}0.50, 0.63{]} & dead \\
22 & pKa, min (PROPKA) \cite{olsson2011,sondergaard2011} & pKa & structure &
0.52 {[}0.50, 0.65{]} & active \\
23 & Amino-acid propensity (Rosetta, site) \cite{alford2017} & Rosetta energy &
structure & 0.52 {[}0.50, 0.59{]} & active \\
24 & Electrostatics (Rosetta, site) \cite{alford2017} & Rosetta energy &
structure & 0.52 {[}0.50, 0.58{]} & active \\
25 & pKa, max (PROPKA) \cite{olsson2011,sondergaard2011} & pKa & structure &
0.52 {[}0.50, 0.67{]} & dead \\
26 & H-bond, backbone to side chain (Rosetta, site) \cite{alford2017} & Rosetta energy &
structure & 0.52 {[}0.50, 0.58{]} & dead \\
\end{longtable}

\textbf{More detail.} Every metric scored on the 21 pairs, 97 in total, with the best-of-158
threshold: Table S5. The wider set of 49 pairs (88 metrics) with the AUROC against the null
distribution of the same state: Table S3. The definitions of the six pair sets and where each is
used: Table S4.

\subsubsection{3.1.2 Distinguishing active and inactive de novo designed enzymes}

\textbf{Dataset.} 192 designs from a deposited design campaign for metallohydrolases
\cite{kim2026}: 16 have a reported kcat/KM (\emph{active}) and 176 do not (here \emph{inactive}: no
kcat/KM has been reported for them, and they have not been confirmed as inactive). The comparison is
that of active against inactive designs. Each design had been selected for testing by the
computational filters of its campaign (predicted active-site geometry and preorganisation)
\cite{kim2026}; 171 of the 192 had been expressed and were soluble, the 16 active designs are the
screening hits characterised with Michaelis--Menten kinetics on purified protein, and the remaining
176 include the 21 that were not expressed as soluble protein \cite{kim2026}. The designs lie in 136
backbone clusters, the 16 active designs in 8 of them (10 of the 16 in two clusters).
Structure-space metrics are computed on the designs as deposited and prediction-based quantities on
Chai-1 \cite{chai2024} predictions with the transition-state-analogue ligand of the design and zinc.
Each item is scored on all 192 designs (Methods 2.2.2).

\textbf{Ranking criteria.} The 31 items are 29 structure-space metrics and 2 prediction-based
metrics of the main set. They are ranked on the direction-free AUROC \cite{hanley1982,mann1947} of
active against inactive designs (0.5 = no separation), with a 95\% bootstrap interval that resamples
the 136 backbone clusters \cite{field2007}. ``Higher in'' is the group in which the metric takes
larger values. A bold AUROC exceeds the best-of-36 threshold \cite{westfall1993}, the 95th
percentile of the best AUROC among 36 unrelated quantities when the activity labels are permuted
over the designs (median 0.66, 95th percentile 0.73); the permutation treats designs as
interchangeable, which underestimates the chance range when the active designs are bundled in few
clusters. AME is measured on the plates against the own model of each design (a different metric
from AME against a deposited structure) and is in Table S6.

\textbf{Result.} Two of the 31 metrics exceed the best-of-36 threshold (0.729): the intra-residue
repulsion (AUROC 0.80) and the repulsion (0.78) of the catalytic residues. The next metrics are the
site total (0.709) and the Dunbrack rotamer term (0.705), both below the threshold. The 192 designs
had passed the filters of their campaign before they were tested, and no kcat/KM has been reported
for the 176 inactive designs.

\begin{longtable}[]{@{}
  >{\raggedright\arraybackslash}p{(\columnwidth - 10\tabcolsep) * \real{0.0709}}
  >{\raggedright\arraybackslash}p{(\columnwidth - 10\tabcolsep) * \real{0.2598}}
  >{\raggedright\arraybackslash}p{(\columnwidth - 10\tabcolsep) * \real{0.1732}}
  >{\raggedright\arraybackslash}p{(\columnwidth - 10\tabcolsep) * \real{0.1181}}
  >{\raggedright\arraybackslash}p{(\columnwidth - 10\tabcolsep) * \real{0.1969}}
  >{\raggedright\arraybackslash}p{(\columnwidth - 10\tabcolsep) * \real{0.1811}}@{}}
\caption*{\textbf{Table 6. Separation of active from inactive designs on the evaluation set of 192
designs (the 25 highest-ranked of 31 metrics).} The 31 metrics are 29 structure-space metrics and 2
prediction-based metrics of the main set, ranked on the direction-free AUROC (lower limit 0.5) of
the 16 active designs against the 176 inactive designs; each row has n = 192 designs (16 active).
Intervals are 95\% bootstrap intervals that resample the 136 backbone clusters. ``Higher in'' is the
group in which the metric takes larger values: ``active'' (the 16 designs with a reported kcat/KM)
or ``inactive'' (the remaining 176 designs, for which no kcat/KM has been reported; they have not
been confirmed as inactive). All 192 designs had passed the filters of their campaign before
testing. A bold AUROC exceeds the best-of-36 threshold, the 95th percentile of the best AUROC among
36 unrelated quantities when the activity labels are permuted over the designs (median 0.66, 95th
percentile 0.73; a reference level for reading the ranking). The six lowest-ranked metrics and every
metric scored on the designs are in Table S6. Source:
main/T6\_plated\_designs\_auroc.csv.}\tabularnewline
\toprule\noalign{}
\hc{Rank} & \hc{Item} & \hc{Group} & \hc{Mode} & \hc{AUROC {[}95\% interval{]}} & \hc{Higher in} \\
\midrule\noalign{}
\endfirsthead
\toprule\noalign{}
\hc{Rank} & \hc{Item} & \hc{Group} & \hc{Mode} & \hc{AUROC {[}95\% interval{]}} & \hc{Higher in} \\
\midrule\noalign{}
\endhead
\bottomrule\noalign{}
\endlastfoot
1 & Intra-residue repulsion (Rosetta, site) \cite{alford2017} & Rosetta energy &
structure & \textbf{0.80 {[}0.64, 0.89{]}} & inactive \\
2 & Repulsion (Rosetta, site) \cite{alford2017} & Rosetta energy &
structure & \textbf{0.78 {[}0.57, 0.92{]}} & inactive \\
3 & Site total (Rosetta) \cite{alford2017} & Rosetta energy &
structure & 0.71 {[}0.52, 0.82{]} & inactive \\
4 & Rotamer, Dunbrack (Rosetta, site) \cite{alford2017} & Rosetta energy &
structure & 0.71 {[}0.58, 0.88{]} & inactive \\
5 & Per-chain pTM minimum (Chai-1) \cite{chai2024,jumper2021} & Interface confidence &
prediction & 0.70 {[}0.54, 0.83{]} & active \\
6 & ipTM (Chai-1) \cite{chai2024,evans2021} & Interface confidence &
prediction & 0.70 {[}0.52, 0.84{]} & active \\
7 & Worst residue (Rosetta, site) \cite{alford2017} & Rosetta energy &
structure & 0.68 {[}0.52, 0.78{]} & inactive \\
8 & Attraction (Rosetta, site) \cite{alford2017} & Rosetta energy &
structure & 0.67 {[}0.57, 0.82{]} & active \\
9 & lk-ball solvation (Rosetta, site) \cite{alford2017} & Rosetta energy &
structure & 0.66 {[}0.54, 0.82{]} & active \\
10 & Total SASA & Geometry & structure & 0.65 {[}0.54, 0.89{]} & active \\
11 & Relative SASA, mean & Geometry &
structure & 0.65 {[}0.53, 0.89{]} & active \\
12 & Electrostatics (Rosetta, site) \cite{alford2017} & Rosetta energy &
structure & 0.64 {[}0.51, 0.79{]} & inactive \\
13 & Pairwise distance, max & Geometry & structure & 0.64 {[}0.51, 0.77{]} & active \\
14 & Relative SASA, max & Geometry &
structure & 0.62 {[}0.52, 0.84{]} & active \\
15 & Pairwise distance, mean & Geometry & structure & 0.61 {[}0.51, 0.77{]} & active \\
16 & Radius of gyration & Geometry & structure & 0.61 {[}0.51, 0.77{]} & active \\
17 & H-bond, side chain (Rosetta, site) \cite{alford2017} & Rosetta energy &
structure & 0.61 {[}0.51, 0.73{]} & inactive \\
18 & Solvation (Rosetta, site) \cite{alford2017} & Rosetta energy &
structure & 0.60 {[}0.51, 0.71{]} & inactive \\
19 & Amino-acid propensity (Rosetta, site) \cite{alford2017} & Rosetta energy &
structure & 0.60 {[}0.51, 0.73{]} & active \\
20 & Polar contacts & Geometry &
structure & 0.59 {[}0.50, 0.74{]} & active \\
21 & Omega torsion (Rosetta, site) \cite{alford2017} & Rosetta energy &
structure & 0.58 {[}0.50, 0.81{]} & active \\
22 & pKa, max (PROPKA) \cite{olsson2011,sondergaard2011} & pKa & structure &
0.56 {[}0.50, 0.78{]} & inactive \\
23 & pKa, range (PROPKA) \cite{olsson2011,sondergaard2011} & pKa & structure &
0.55 {[}0.50, 0.77{]} & inactive \\
24 & Net van der Waals (Rosetta, site) \cite{alford2017} & Rosetta energy &
structure & 0.54 {[}0.50, 0.80{]} & active \\
25 & Fraction buried & Geometry &
structure & 0.53 {[}0.50, 0.70{]} & inactive \\
\end{longtable}

\textbf{More detail.} The six lowest-ranked metrics and every other metric scored on the 192 designs
(71 in total), with the best-of-128 threshold: Table S6. The ranking of the metrics among the 16
active designs is in 3.3.2.

\begin{center}\rule{0.5\linewidth}{0.5pt}\end{center}

\subsection{3.2 Substrate discrimination test}

\textbf{Experiment.} Each enzyme is predicted with Chai-1 \cite{chai2024} in four conditions that
differ only in the ligand presented: the cognate ligand, a wrong ligand of the same enzyme class, a
wrong ligand of a different class, and a decoy ligand taken from an unrelated structure. Each
prediction-based quantity is ranked by how well it separates the cognate prediction from a
prediction with a wrong ligand for the same enzyme (Methods 2.3).

\textbf{Dataset.} 55 natural enzymes from M-CSA \cite{ribeiro2018}, of the 59 in the experiment:
for these, every scored quantity has a defined value in all four conditions and every ligand
presented was present in the prediction (four enzymes are excluded because a haem-type ligand could
not be built as a chemical structure and the predictor dropped it; the analysis on all 59 is in
Table S7). One n, 55 enzymes, applies to every row (Methods 2.3).

\textbf{Ranking criteria.} Chai-1 returns five models per prediction, and each score is summarised
over these models as the mean, maximum or spread (standard deviation over the models), and for the
AME RMSD also as the minimum, so that one metric appears in several rows. The 17 quantities listed
are the four prediction-based metrics of the main set (ipTM, per-chain pTM minimum, ligand
clearance, AME RMSD), the AME pass bit and four bookkeeping counters. Each quantity is ranked on the
mean of three direction-free AUROCs \cite{hanley1982,mann1947}, one per kind of wrong ligand (same
class, different class, decoy), each with a 95\% interval that resamples enzymes. ``Higher in'' is
the condition in which the quantity takes larger values. A bold mean exceeds the best-of-30
threshold \cite{westfall1993}, the 95th percentile of the best AUROC among 30 quantities analysed
when the four ligand conditions are permuted within each enzyme (median 0.56, 95th percentile 0.61).

\begin{longtable}[]{@{}
  >{\raggedright\arraybackslash}p{(\columnwidth - 16\tabcolsep) * \real{0.0573}}
  >{\raggedright\arraybackslash}p{(\columnwidth - 16\tabcolsep) * \real{0.1656}}
  >{\raggedright\arraybackslash}p{(\columnwidth - 16\tabcolsep) * \real{0.1019}}
  >{\raggedright\arraybackslash}p{(\columnwidth - 16\tabcolsep) * \real{0.1529}}
  >{\raggedright\arraybackslash}p{(\columnwidth - 16\tabcolsep) * \real{0.1274}}
  >{\raggedright\arraybackslash}p{(\columnwidth - 16\tabcolsep) * \real{0.1274}}
  >{\raggedright\arraybackslash}p{(\columnwidth - 16\tabcolsep) * \real{0.1274}}
  >{\raggedright\arraybackslash}p{(\columnwidth - 16\tabcolsep) * \real{0.0573}}
  >{\raggedright\arraybackslash}p{(\columnwidth - 16\tabcolsep) * \real{0.0828}}@{}}
\caption*{\textbf{Table 7. The substrate swap on the evaluation set of 55 enzymes, 17 scored
prediction-based quantities ranked by AUROC.} Direction-free AUROC (lower limit 0.5) of the cognate
prediction against a prediction with a wrong ligand of the same enzyme, in three columns by kind of
wrong ligand, with 95\% intervals that resample enzymes; the value of an enzyme in a condition is
the mean over the seeds 101, 202 and 303. Rows are ranked by the mean of the three AUROCs; a bold
mean exceeds the best-of-30 threshold (median 0.56, 95th percentile 0.61; permutation of the four
ligand conditions within each enzyme over 30 quantities analysed, defined in the Methods; a
reference level for reading the ranking). Chai-1 predicts five models for each prediction, and each
score is summarised over these models (mean, maximum, spread as the standard deviation over the
models, and for the AME RMSD also the minimum), so that one metric appears in several rows, one per
summary. Source: main/T7\_substrate\_swap.csv.}\tabularnewline
\toprule\noalign{}
\hc{Rank} & \hc{Quantity} & \hc{Summary over the five models} & \hc{Group} & \hc{Same class} &
\hc{Different class} & \hc{Decoy} & \hc{Mean} & \hc{Higher in} \\
\midrule\noalign{}
\endfirsthead
\toprule\noalign{}
\hc{Rank} & \hc{Quantity} & \hc{Summary over the five models} & \hc{Group} & \hc{Same class} &
\hc{Different class} & \hc{Decoy} & \hc{Mean} & \hc{Higher in} \\
\midrule\noalign{}
\endhead
\bottomrule\noalign{}
\endlastfoot
1 & ipTM \cite{chai2024,evans2021} & mean & Interface confidence (main set) & 0.69
{[}0.60, 0.78{]} & 0.75 {[}0.66, 0.84{]} &
0.65 {[}0.55, 0.76{]} & \textbf{0.70} & cognate \\
2 & ipTM \cite{chai2024,evans2021} & maximum &
Interface confidence (main set) & 0.68 {[}0.59, 0.77{]} &
0.74 {[}0.65, 0.83{]} & 0.65 {[}0.54, 0.74{]} & \textbf{0.69} & cognate \\
3 & ipTM \cite{chai2024,evans2021} & spread (SD) &
Interface confidence (main set) & 0.70 {[}0.60, 0.79{]} &
0.69 {[}0.60, 0.77{]} & 0.67 {[}0.59, 0.75{]} & \textbf{0.69} & wrong ligand \\
4 & Ligand clearance \cite{ahern2026} & best model &
Active-site accuracy (main set) & 0.61 {[}0.52, 0.69{]} &
0.65 {[}0.57, 0.74{]} & 0.62 {[}0.53, 0.71{]} & \textbf{0.63} & wrong ligand \\
5 & Per-chain pTM minimum \cite{chai2024,jumper2021} & spread (SD) &
Interface confidence (main set) & 0.58 {[}0.51, 0.68{]} &
0.58 {[}0.51, 0.68{]} & 0.69 {[}0.59, 0.79{]} & \textbf{0.62} & wrong ligand \\
6 & AME RMSD \cite{ahern2026} & spread (SD) &
Active-site accuracy (main set) & 0.55 {[}0.51, 0.60{]} &
0.56 {[}0.51, 0.62{]} & 0.57 {[}0.52, 0.63{]} & 0.56 & wrong ligand \\
7 & Per-chain pTM minimum \cite{chai2024,jumper2021} & mean & Interface confidence
(main set) & 0.58 {[}0.51, 0.67{]} & 0.60 {[}0.51, 0.71{]} &
0.50 {[}0.50, 0.62{]} & 0.56 & cognate \\
8 & Per-chain pTM minimum \cite{chai2024,jumper2021} & maximum &
Interface confidence (main set) & 0.58 {[}0.51, 0.67{]} &
0.59 {[}0.51, 0.71{]} & 0.50 {[}0.50, 0.62{]} & 0.56 & cognate \\
9 & AME RMSD \cite{ahern2026} & maximum &
Active-site accuracy (main set) & 0.53 {[}0.50, 0.57{]} &
0.54 {[}0.51, 0.58{]} & 0.54 {[}0.51, 0.59{]} & 0.54 & wrong ligand \\
10 & AME RMSD \cite{ahern2026} & minimum &
Active-site accuracy (main set) & 0.53 {[}0.50, 0.56{]} &
0.54 {[}0.51, 0.57{]} & 0.52 {[}0.50, 0.57{]} & 0.53 & wrong ligand \\
11 & AME RMSD, no symmetry resolution \cite{ahern2026} & minimum &
Active-site accuracy (main set) & 0.52 {[}0.50, 0.56{]} &
0.54 {[}0.51, 0.57{]} & 0.53 {[}0.50, 0.57{]} & 0.53 & wrong ligand \\
12 & AME RMSD \cite{ahern2026} & mean &
Active-site accuracy (main set) & 0.53 {[}0.50, 0.56{]} & 0.53
{[}0.51, 0.57{]} & 0.52 {[}0.50, 0.57{]} & 0.53 & wrong ligand \\
13 & Symmetry swaps (count) & single value & Bookkeeping counter & 0.51 {[}0.50, 0.54{]} &
0.51 {[}0.50, 0.53{]} & 0.50 {[}0.50, 0.52{]} & 0.51 & cognate \\
14 & AME pass bit \cite{ahern2026} & single value &
Active-site accuracy (pass bit) & 0.50 {[}0.50, 0.50{]} &
0.50 {[}0.50, 0.53{]} & 0.50 {[}0.50, 0.53{]} & 0.50 & mixed \\
15 & Backbone atoms aligned (count) & single value &
Bookkeeping counter & 0.50 {[}0.50, 0.50{]} & 0.50 {[}0.50, 0.50{]} &
0.50 {[}0.50, 0.50{]} & 0.50 & mixed \\
16 & Catalytic atoms (count) & single value & Bookkeeping counter & 0.50 {[}0.50, 0.50{]} &
0.50 {[}0.50, 0.50{]} & 0.50 {[}0.50, 0.50{]} & 0.50 & mixed \\
17 & Models predicted (count) & single value &
Bookkeeping counter & 0.50 {[}0.50, 0.50{]} & 0.50 {[}0.50, 0.50{]} &
0.50 {[}0.50, 0.50{]} & 0.50 & mixed \\
\end{longtable}

\textbf{More detail.} The analysis on all 59 natural enzymes (mean over five seeds of the cognate
prediction) and on the 30 de novo designs, for the scored prediction-based quantities, including the
combined Chai-1 score, with the paired change in units of seed noise: Table S7. The same comparison
stratified by the difference in size and charge of the ligand, with paired win fractions: Table S8.

\begin{center}\rule{0.5\linewidth}{0.5pt}\end{center}

\subsection{3.3 Activity ranking test}

\textbf{Experiment.} Metrics are scored on real proteins whose activity has been measured, and each
metric is ranked by the rank correlation between its value and the measured activity (Methods 2.4).
Two datasets are used, each with its own table.

\subsubsection{3.3.1 Ranking the activity of natural enzymes}

\textbf{Dataset.} BglB, a β-glucosidase from \emph{Paenibacillus polymyxa}, with single-point
variants characterised by the D2DCure consortium \cite{carlin2016} (data retrieved from d2dcure.com
on 2026-08-06): 655 variants at 207 positions, of which the \textbf{432 variants at 175 positions
that are folded (expressed and not destabilised) and have kinetics} form the evaluation set. The
measured value is the \emph{impairment}, the log10 of the loss of kcat/KM relative to the wild type
(larger = more strongly impaired). Each variant is built in the computer on the crystal structure of
the wild type (RCSB PDB entry 2JIE \cite{burley2025,berman2000}, which contains a covalent glucosyl
intermediate) and structure-space metrics are computed over the two catalytic glutamates (Glu167 and
Glu356, from the active-site features in UniProt \cite{uniprot2023}) before and after the
substitution; the variants are single-point substitutions of one metal-independent glycosidase, and
ranking them is a different task from ranking de novo designs (3.3.2); prediction-based metrics are
computed on Chai-1 \cite{chai2024} predictions of the variant with the assay substrate (pNPG)
(Methods 2.4.1).

\textbf{Ranking criteria.} The 32 items are the 28 structure-space metrics of the main set that
BglB defines (the omega term does not change with a side-chain substitution on a fixed backbone) and
the 4 prediction-based metrics of the main set. Each item is ranked on the Spearman rank correlation
ρ \cite{spearman1904} with the measured impairment, with a 95\% bootstrap interval that resamples the
175 positions \cite{field2007}. The score is the size of the change of the metric (the absolute
difference between the variant and the unsubstituted protein), with the direction defined as larger
in more strongly impaired variants. A bold ρ exceeds the best-of-41 threshold \cite{westfall1993},
the 95th percentile of the best ρ among 41 unrelated quantities when the measured impairments are
permuted over whole positions (median 0.24, 95th percentile 0.32).

\textbf{Result.} 24 of the 32 metrics exceed the best-of-41 threshold (0.32). The highest ρ is 0.48,
that of the pKa range of the catalytic residues. The ρ values of these 24 metrics lie between 0.33
and 0.48, and their 95\% intervals all contain the range from 0.35 to 0.45.

\begin{longtable}[]{@{}
  >{\raggedright\arraybackslash}p{(\columnwidth - 8\tabcolsep) * \real{0.0818}}
  >{\raggedright\arraybackslash}p{(\columnwidth - 8\tabcolsep) * \real{0.3364}}
  >{\raggedright\arraybackslash}p{(\columnwidth - 8\tabcolsep) * \real{0.2091}}
  >{\raggedright\arraybackslash}p{(\columnwidth - 8\tabcolsep) * \real{0.1364}}
  >{\raggedright\arraybackslash}p{(\columnwidth - 8\tabcolsep) * \real{0.2364}}@{}}
\caption*{\textbf{Table 8. BglB variants: the 26 highest-ranked of 32 metrics on the evaluation set
of 432 variants.} Spearman rank correlation ρ between each metric and the measured impairment of
the 432 variants, with a 95\% interval that resamples the 175 positions. The score is the size of
the change of the metric, with the direction defined as a larger change in more strongly impaired
variants; a negative ρ is reported as such. The 32 metrics are the metrics of the main set that
BglB defines. A value in bold exceeds the best-of-41 threshold (median 0.24, 95th percentile 0.32;
permutation of the measured impairments over whole positions; a reference level for reading the
ranking). Rows with structure-space metrics have 431 variants with a value and rows with
prediction-based metrics 432. The six lowest-ranked metrics and every quantity analysed are in
Table S9. Source: main/T8\_bglb\_variants.csv.}\tabularnewline
\toprule\noalign{}
\hc{Rank} & \hc{Item} & \hc{Group} & \hc{Mode} & \hc{Spearman ρ {[}95\% interval{]}} \\
\midrule\noalign{}
\endfirsthead
\toprule\noalign{}
\hc{Rank} & \hc{Item} & \hc{Group} & \hc{Mode} & \hc{Spearman ρ {[}95\% interval{]}} \\
\midrule\noalign{}
\endhead
\bottomrule\noalign{}
\endlastfoot
1 & pKa, range (PROPKA) \cite{olsson2011,sondergaard2011} & pKa & structure &
\textbf{0.48 {[}0.35, 0.59{]}} \\
2 & pKa, min (PROPKA) \cite{olsson2011,sondergaard2011} & pKa & structure &
\textbf{0.46 {[}0.32, 0.57{]}} \\
3 & pKa shift, max absolute (PROPKA) \cite{olsson2011,sondergaard2011} & pKa & structure &
\textbf{0.46 {[}0.32, 0.58{]}} \\
4 & pKa, max (PROPKA) \cite{olsson2011,sondergaard2011} & pKa & structure &
\textbf{0.46 {[}0.32, 0.58{]}} \\
5 & Intra-residue repulsion (Rosetta, site) \cite{alford2017} & Rosetta energy &
structure & \textbf{0.45 {[}0.30, 0.57{]}} \\
6 & Relative SASA, mean & Geometry & structure & \textbf{0.45 {[}0.32, 0.57{]}} \\
7 & pKa shift, mean (PROPKA) \cite{olsson2011,sondergaard2011} & pKa & structure &
\textbf{0.45 {[}0.31, 0.57{]}} \\
8 & pKa, mean (PROPKA) \cite{olsson2011,sondergaard2011} & pKa & structure &
\textbf{0.45 {[}0.31, 0.57{]}} \\
9 & Total SASA & Geometry & structure & \textbf{0.45 {[}0.31, 0.57{]}} \\
10 & Repulsion (Rosetta, site) \cite{alford2017} & Rosetta energy &
structure & \textbf{0.45 {[}0.31, 0.57{]}} \\
11 & Pairwise distance, max & Geometry & structure & \textbf{0.45 {[}0.30, 0.56{]}} \\
12 & Attraction (Rosetta, site) \cite{alford2017} & Rosetta energy &
structure & \textbf{0.44 {[}0.30, 0.56{]}} \\
13 & Rotamer, Dunbrack (Rosetta, site) \cite{alford2017} & Rosetta energy &
structure & \textbf{0.44 {[}0.29, 0.56{]}} \\
14 & lk-ball solvation (Rosetta, site) \cite{alford2017} & Rosetta energy &
structure & \textbf{0.44 {[}0.31, 0.54{]}} \\
15 & Radius of gyration & Geometry & structure & \textbf{0.43 {[}0.28, 0.56{]}} \\
16 & Net van der Waals (Rosetta, site) \cite{alford2017} & Rosetta energy &
structure & \textbf{0.43 {[}0.28, 0.55{]}} \\
17 & Relative SASA, max & Geometry & structure & \textbf{0.43 {[}0.29, 0.54{]}} \\
18 & Pairwise distance, mean & Geometry & structure & \textbf{0.43 {[}0.28, 0.55{]}} \\
19 & Electrostatics (Rosetta, site) \cite{alford2017} & Rosetta energy &
structure & \textbf{0.43 {[}0.28, 0.55{]}} \\
20 & Site total (Rosetta) \cite{alford2017} & Rosetta energy &
structure & \textbf{0.42 {[}0.27, 0.55{]}} \\
21 & Worst residue (Rosetta, site) \cite{alford2017} & Rosetta energy &
structure & \textbf{0.42 {[}0.26, 0.55{]}} \\
22 & Solvation (Rosetta, site) \cite{alford2017} & Rosetta energy &
structure & \textbf{0.41 {[}0.27, 0.54{]}} \\
23 & H-bond, side chain (Rosetta, site) \cite{alford2017} & Rosetta energy &
structure & \textbf{0.35 {[}0.19, 0.49{]}} \\
24 & ipTM (Chai-1) \cite{chai2024,evans2021} & Interface confidence &
prediction & \textbf{0.33 {[}0.19, 0.45{]}} \\
25 & Per-chain pTM minimum (Chai-1) \cite{chai2024,jumper2021} & Interface confidence &
prediction & 0.28 {[}0.15, 0.41{]} \\
26 & Fraction buried & Geometry & structure & 0.19 {[}0.06, 0.30{]} \\
\end{longtable}

\textbf{More detail.} Every quantity analysed, ranked, with the AUROC of the binary contrast
(impaired against wild-type-like) alongside each rank correlation: Table S9. The 39 structure-space
metrics with the signed, direction-free and size-based AUROC: Table S10. The 15 prediction-based
metrics, with the signed correlation of the change: Table S11.

\subsubsection{3.3.2 Ranking the activity of the 16 de novo designs with a reported kcat/KM}

\textbf{Dataset.} The ranking is made on 16 designs, not on the 192 of the campaign of 3.1.2
\cite{kim2026}: only the 16 designs with a reported kcat/KM have a measured activity, they form the
evaluation set, and each row has n = 16. The other 176 designs have no measured activity and are
not ranked. The 16 lie in 8 of the 136 backbone clusters, 10 of them in two clusters (Methods
2.4.2). They are the screening hits characterised with Michaelis--Menten kinetics on purified
protein, five from the first campaign and eleven from the second; the reported kcat/KM is 35 to 140
M\textsuperscript{−1} s\textsuperscript{−1} for four of the five hits from the first campaign,
16,000 M\textsuperscript{−1} s\textsuperscript{−1} for the fifth, and more than 1,000
M\textsuperscript{−1} s\textsuperscript{−1} for all eleven hits from the second campaign
\cite{kim2026}.

\textbf{Ranking criteria.} The 31 items are those of Table 6. Each item is ranked on the size of the
Spearman correlation ρ \cite{spearman1904} between its value on the 16 designs as deposited and
their log10 kcat/KM, with a 90\% bootstrap interval over the 16 designs. The sign of ρ is reported
(``higher value means higher or lower activity''). A bold ρ has an interval that excludes 0.

\textbf{Result.} On the 16 designs, four of the 31 items have a 90\% interval that excludes 0: polar
contacts (ρ = 0.67), net van der Waals (−0.63), solvation (0.57) and amino-acid propensity (0.54).
The intervals resample the 16 designs, which lie in 8 backbone clusters. The analysis of hits per
plate is separate from this ranking: it uses the active and inactive labels of all 192 designs and
fills a plate with the quarter of the designs with the highest, or with the lowest, values of an
item. An unfiltered plate contains 8.0 hits and 95\% of random choices contain at most 14.0; 4 of
the 31 items exceed 14.0 at one end, with 22 hits for the intra-residue repulsion and for the
repulsion, and 16 for the maximum pairwise distance and for the Dunbrack rotamer term (Table S12).

\begin{longtable}[]{@{}
  >{\raggedright\arraybackslash}p{(\columnwidth - 10\tabcolsep) * \real{0.0738}}
  >{\raggedright\arraybackslash}p{(\columnwidth - 10\tabcolsep) * \real{0.2787}}
  >{\raggedright\arraybackslash}p{(\columnwidth - 10\tabcolsep) * \real{0.1639}}
  >{\raggedright\arraybackslash}p{(\columnwidth - 10\tabcolsep) * \real{0.1148}}
  >{\raggedright\arraybackslash}p{(\columnwidth - 10\tabcolsep) * \real{0.2131}}
  >{\raggedright\arraybackslash}p{(\columnwidth - 10\tabcolsep) * \real{0.1557}}@{}}
\caption*{\textbf{Table 9. Ranking of the metrics on the 16 designs with a reported kcat/KM: the 25
highest-ranked of 31.} Spearman rank correlation ρ between each metric (original value on the
design) and log10 kcat/KM over the 16 designs, with a 90\% interval that resamples the 16 designs;
metrics are ranked on the size of ρ. The other 176 of the 192 designs have no measured activity and
are not part of this ranking. A value in bold has an interval that excludes 0. The 31 metrics are 29
structure-space metrics and 2 prediction-based metrics of the main set. The 16 screening hits with a
reported kcat/KM lie in 8 backbone clusters. The six lowest-ranked metrics, and every metric with a
correlation, are in Table S13; the analysis of hits per plate, which uses the active and inactive
labels of the 192 designs, is in Table S12. Source:
main/T9\_metric\_ranking\_16\_designs.csv.}\tabularnewline
\toprule\noalign{}
\hc{Rank} & \hc{Item} & \hc{Group} & \hc{Mode} & \hc{Spearman ρ {[}90\% interval{]}} &
\hc{Higher value means} \\
\midrule\noalign{}
\endfirsthead
\toprule\noalign{}
\hc{Rank} & \hc{Item} & \hc{Group} & \hc{Mode} & \hc{Spearman ρ {[}90\% interval{]}} &
\hc{Higher value means} \\
\midrule\noalign{}
\endhead
\bottomrule\noalign{}
\endlastfoot
1 & Polar contacts & Geometry & structure & \textbf{0.67 {[}0.41, 0.85{]}} & higher activity \\
2 & Net van der Waals (Rosetta, site) \cite{alford2017} & Rosetta energy &
structure & \textbf{−0.63 {[}−0.88, −0.24{]}} & lower activity \\
3 & Solvation (Rosetta, site) \cite{alford2017} & Rosetta energy &
structure & \textbf{0.57 {[}0.15, 0.85{]}} & higher activity \\
4 & Amino-acid propensity (Rosetta, site) \cite{alford2017} & Rosetta energy &
structure & \textbf{0.54 {[}0.08, 0.81{]}} & higher activity \\
5 & Pairwise distance, mean & Geometry & structure & −0.43 {[}−0.71, 0.02{]} & lower activity \\
6 & Site total (Rosetta) \cite{alford2017} & Rosetta energy &
structure & −0.42 {[}−0.69, 0.01{]} & lower activity \\
7 & Radius of gyration & Geometry & structure & −0.41 {[}−0.69, 0.03{]} & lower activity \\
8 & H-bond, side chain (Rosetta, site) \cite{alford2017} & Rosetta energy &
structure & −0.40 {[}−0.74, 0.05{]} & lower activity \\
9 & Ramachandran (Rosetta, site) \cite{alford2017} & Rosetta energy &
structure & −0.38 {[}−0.82, 0.15{]} & lower activity \\
10 & Repulsion (Rosetta, site) \cite{alford2017} & Rosetta energy &
structure & −0.33 {[}−0.73, 0.17{]} & lower activity \\
11 & pKa, mean (PROPKA) \cite{olsson2011,sondergaard2011} & pKa & structure &
−0.30 {[}−0.57, 0.09{]} & lower activity \\
12 & pKa shift, mean (PROPKA) \cite{olsson2011,sondergaard2011} & pKa & structure &
−0.30 {[}−0.56, 0.09{]} & lower activity \\
13 & Relative SASA, mean & Geometry & structure & −0.25 {[}−0.76, 0.31{]} & lower activity \\
14 & Total SASA & Geometry & structure & −0.25 {[}−0.76, 0.30{]} & lower activity \\
15 & Relative SASA, max & Geometry & structure & −0.25 {[}−0.76, 0.28{]} & lower activity \\
16 & Attraction (Rosetta, site) \cite{alford2017} & Rosetta energy &
structure & −0.25 {[}−0.66, 0.25{]} & lower activity \\
17 & Pairwise distance, max & Geometry & structure & 0.23 {[}−0.26, 0.64{]} & higher activity \\
18 & Intra-residue repulsion (Rosetta, site) \cite{alford2017} & Rosetta energy &
structure & −0.23 {[}−0.65, 0.30{]} & lower activity \\
19 & lk-ball solvation (Rosetta, site) \cite{alford2017} & Rosetta energy &
structure & −0.21 {[}−0.53, 0.19{]} & lower activity \\
20 & Electrostatics (Rosetta, site) \cite{alford2017} & Rosetta energy &
structure & −0.19 {[}−0.58, 0.25{]} & lower activity \\
21 & Per-chain pTM minimum (Chai-1) \cite{chai2024,jumper2021} & Interface confidence &
prediction & 0.18 {[}−0.35, 0.66{]} & higher activity \\
22 & Rotamer, Dunbrack (Rosetta, site) \cite{alford2017} & Rosetta energy &
structure & 0.18 {[}−0.31, 0.57{]} & higher activity \\
23 & H-bond, backbone to side chain (Rosetta, site) \cite{alford2017} & Rosetta energy &
structure & 0.17 {[}−0.21, 0.52{]} & higher activity \\
24 & pKa, max (PROPKA) \cite{olsson2011,sondergaard2011} & pKa & structure &
−0.16 {[}−0.49, 0.26{]} & lower activity \\
25 & pKa, range (PROPKA) \cite{olsson2011,sondergaard2011} & pKa & structure &
−0.13 {[}−0.53, 0.31{]} & lower activity \\
\end{longtable}

\textbf{More detail.} The correlation with kcat/KM of every metric that has one: Table S13. The hits
per plate of every metric when it fills a 96-well plate with the quarter of the 192 designs with its
highest, or with its lowest, values (active against inactive; the designs are not ranked by
activity): Table S12. The enrichment at keep fractions of 5, 10, 25 and 50\% in both directions:
Table S14 (55 structure-space metrics) and Table S15 (16 prediction-based metrics).

\begin{center}\rule{0.5\linewidth}{0.5pt}\end{center}

\subsection{3.4 Detection ability test}

\textbf{Experiment.} Each metric is ranked on whether it changes more under a lesion at a catalytic
residue than under the same lesion at a matched non-catalytic control position of the same protein
(Methods 2.5). Two kinds of lesion are tested, each with its own datasets: the catalytic lesion
(3.4.1, Table 10) and the second-shell lesion (3.4.2, Table S28).

\subsubsection{3.4.1 Detection of the catalytic lesion}

\textbf{Experiment.} The catalytic residues are replaced in four steps of increasing change in
side-chain volume (isosteric 15 Å³, non-isosteric 36 Å³, to Ala 53 Å³, to Gly 81 Å³), all at once in
the crystal structure by an operation in rotamer space and a repacking, or one residue per
prediction in the sequence of the re-predicted complex with Chai-1 \cite{chai2024}; the control makes
the same substitutions at matched buried non-catalytic positions of the same protein.

\textbf{Dataset.} For the 29 structure-space metrics of the main set: the 143 M-CSA enzymes
\cite{ribeiro2018} (77 enzyme sub-subclasses; packing-matched, with count gating) pooled with the 52
enzymes of the set of 53 enzymes that are not in it (active sites from UniProt \cite{uniprot2023} and
M-CSA; controls matched only on residue type and burial; the one enzyme of that set that is the same
protein as an enzyme of the main set is counted once), 195 enzymes in total. For the four
prediction-based metrics of the main set: the natural arm of the re-predicted ladder, the 59 natural
enzymes of the substrate-swap set with 286 catalytic/control pairs. The 30 de novo designs and the
15 designed systems of the ladder, whose control is constructed, are analysed separately (the
designs form the de novo sensitivity set of Table S20) (Methods 2.5.1).

\textbf{Ranking criteria.} The 33 items are the 29 structure-space metrics and the 4
prediction-based metrics of the main set. The rank statistic R is the probability that a metric
changes more at the catalytic lesion than at the matched control of the same protein (0.5 = no
specificity; above 0.5, more change at the catalytic lesion). The change of a metric is the absolute
difference between the perturbed and the unperturbed structure (or prediction), a tie counts for
half, and proteins in which neither arm moves are omitted. A step counts for a structure-space
metric only if the metric responds to the lesion (the interval of its median change excludes zero)
and the control response lies above the metric's own noise; a step that does not count is in
parentheses, and ``--'' is a step with fewer than five informative proteins. The mean R over the
counted steps has a 95\% bootstrap interval that resamples enzyme sub-subclasses (systems for the
predictor). The rows are grouped by the lesion at which the metric is first detected (the smallest
step at which it responds; ``Never'' if at none) and ranked within each group by mean R; a metric
without a counted step has no mean and closes its group. A bold mean exceeds the best-of-33
threshold \cite{westfall1993}, the 95th percentile of the best R among 33 quantities when the
catalytic and control labels are swapped within each protein (median 0.57, 95th percentile 0.62).
The number of counted steps per metric and the smallest step at which a structure-space metric is
specific (meeting the five conditions of Methods 2.5.1) are in Table S19.

\textbf{Result.} Six of the 33 metrics have a mean R above the best-of-33 threshold (0.62): the
per-chain pTM minimum (0.75), the AME RMSD (0.68), the catalytic pKa range (0.68), ipTM (0.67), the
fraction buried (0.66) and the solvation of the catalytic residues (0.63).

\emph{Counted steps differ between metrics.} The mean R is taken over the counted steps only, and
the number of counted steps differs between metrics (last column of Table 10): 16 of the 33 metrics
have four, 6 have three, 6 have two, 3 have one and 2 have none. A mean over two steps (the pKa
range, over the isosteric and non-isosteric steps; the fraction buried, over the Ala and Gly steps)
is therefore taken over a different set of steps than a mean over four, and the rows of Table 10 are
not strictly comparable.

\emph{Metrics without a mean.} Polar contacts have an R of 0.93 to 0.97 at all four steps (101 to
115 enzymes in which at least one arm moves) and respond to the lesion at every step, but their
control response lies above the metric's noise at no step, where R measures whether the metric moves
at all and not whether it is specific (Methods 2.5.1), so no step is counted and there is no mean.
The maximum pairwise distance (R 0.19 to 0.36 at three steps) responds to the lesion at no step and
likewise has no mean.

\emph{Metrics that changed more at the control.} Four metrics have a mean R whose 95\% interval lies
below 0.5, so that they changed more at the matched control than at the catalytic lesion:
electrostatics (0.24 {[}0.18, 0.30{]}), the backbone-to-side-chain hydrogen bond (0.29
{[}0.19, 0.39{]}), the site total (0.42 {[}0.37, 0.47{]}) and the mean pairwise distance (0.41
{[}0.34, 0.50{]}; upper limit 0.496).

\begin{longtable}[]{@{}
  >{\raggedright\arraybackslash}p{(\columnwidth - 10\tabcolsep) * \real{0.1138}}
  >{\raggedright\arraybackslash}p{(\columnwidth - 10\tabcolsep) * \real{0.0732}}
  >{\raggedright\arraybackslash}p{(\columnwidth - 10\tabcolsep) * \real{0.2764}}
  >{\raggedright\arraybackslash}p{(\columnwidth - 10\tabcolsep) * \real{0.2439}}
  >{\raggedright\arraybackslash}p{(\columnwidth - 10\tabcolsep) * \real{0.1951}}
  >{\raggedright\arraybackslash}p{(\columnwidth - 10\tabcolsep) * \real{0.0976}}@{}}
\caption*{\textbf{Table 10. Catalytic lesion (chemistry ladder): the 33 ranked metrics grouped by
the lesion at which they are first detected and ranked by R within the group.} Rank statistic R,
the probability that a metric changes more at the catalytic lesion than at the matched control
(0.5 = no specificity), at the four steps and as the mean over the counted steps, with a 95\%
interval that resamples enzyme sub-subclasses (systems for the predictor). ``Detected from'' is the
smallest lesion step (isosteric 15 Å³, non-isosteric 36 Å³, Ala 53 Å³, Gly 81 Å³) at which the
metric responds. Structure-space metrics: up to 195 natural enzymes (143 + 52 from the 53-enzyme
set; the number of enzymes in which a metric moves at all ranges from 0 to 150 per metric and step);
prediction-based metrics (ipTM, per-chain pTM minimum, ligand clearance, AME RMSD): 59 natural
enzymes, 286 pairs. A step in parentheses is not counted; the mean R is taken over the counted steps
only, whose number (last column) differs between metrics, so that means over different sets of steps
are not strictly comparable. A mean in bold exceeds the best-of-33 threshold (median 0.57, 95th
percentile 0.62; a reference level for reading the ranking). The 8 other rows (6 supplementary
metrics and 2 aliases of ranked metrics), the number of counted steps, the onset of specificity and
all per-step flags are in Table S19. Source:
main/T10\_specificity\_catalytic\_lesion.csv.}\tabularnewline
\toprule\noalign{}
\hc{Detected from} & \hc{Rank in group} & \hc{Item} & \hc{R at isosteric / non-isosteric / Ala / Gly} &
\hc{Mean R {[}95\% interval{]}} & \hc{Counted steps (of 4)} \\
\midrule\noalign{}
\endfirsthead
\toprule\noalign{}
\hc{Detected from} & \hc{Rank in group} & \hc{Item} & \hc{R at isosteric / non-isosteric / Ala / Gly} &
\hc{Mean R {[}95\% interval{]}} & \hc{Counted steps (of 4)} \\
\midrule\noalign{}
\endhead
\bottomrule\noalign{}
\endlastfoot
Isosteric & 1 & Per-chain pTM minimum (Chai-1) \cite{chai2024,jumper2021} &
0.81 / 0.73 / 0.75 / 0.72 & \textbf{0.75 {[}0.68, 0.82{]}} & 4 \\
Isosteric & 2 & pKa, range (PROPKA) \cite{olsson2011,sondergaard2011} &
0.70 / 0.67 / -- / -- & \textbf{0.68 {[}0.58, 0.78{]}} & 2 \\
Isosteric & 3 & ipTM (Chai-1) \cite{chai2024,evans2021} &
0.71 / 0.71 / 0.69 / 0.58 & \textbf{0.67 {[}0.60, 0.75{]}} & 4 \\
Isosteric & 4 & Solvation (Rosetta, site) \cite{alford2017} &
0.59 / 0.73 / 0.60 / 0.62 & \textbf{0.63 {[}0.57, 0.70{]}} & 4 \\
Isosteric & 5 & Relative SASA, mean & 0.48 / 0.55 / 0.73 / 0.69 & 0.61 {[}0.56, 0.67{]} & 4 \\
Isosteric & 6 & pKa, min (PROPKA) \cite{olsson2011,sondergaard2011} &
0.54 / 0.67 / -- / -- & 0.61 {[}0.53, 0.67{]} & 2 \\
Isosteric & 7 & Repulsion (Rosetta, site) \cite{alford2017} &
0.48 / 0.55 / 0.64 / 0.62 & 0.57 {[}0.52, 0.62{]} & 4 \\
Isosteric & 8 & pKa shift, mean (PROPKA) \cite{olsson2011,sondergaard2011} &
0.46 / 0.67 / -- / -- & 0.56 {[}0.47, 0.66{]} & 2 \\
Isosteric & 9 & Relative SASA, max & 0.49 / (0.50) / 0.56 / 0.55 & 0.53 {[}0.48, 0.59{]} & 3 \\
Isosteric & 10 & Attraction (Rosetta, site) \cite{alford2017} &
0.53 / 0.58 / 0.55 / 0.46 & 0.53 {[}0.46, 0.60{]} & 4 \\
Isosteric & 11 & Ramachandran (Rosetta, site) \cite{alford2017} &
0.43 / 0.55 / 0.55 / 0.56 & 0.52 {[}0.47, 0.59{]} & 4 \\
Isosteric & 12 & Intra-residue repulsion (Rosetta, site) \cite{alford2017} &
0.48 / 0.59 / 0.48 / 0.52 & 0.52 {[}0.47, 0.57{]} & 4 \\
Isosteric & 13 & Rotamer, Dunbrack (Rosetta, site) \cite{alford2017} &
0.45 / 0.47 / 0.58 / 0.58 & 0.52 {[}0.46, 0.57{]} & 4 \\
Isosteric & 14 & Radius of gyration & 0.42 / 0.48 / 0.62 / -- & 0.51 {[}0.45, 0.57{]} & 3 \\
Isosteric & 15 & Amino-acid propensity (Rosetta, site) \cite{alford2017} &
0.45 / (0.57) / 0.52 / 0.53 & 0.50 {[}0.45, 0.55{]} & 3 \\
Isosteric & 16 & pKa, max (PROPKA) \cite{olsson2011,sondergaard2011} &
0.43 / 0.56 / -- / -- & 0.49 {[}0.39, 0.61{]} & 2 \\
Isosteric & 17 & Worst residue (Rosetta, site) \cite{alford2017} &
0.45 / 0.46 / 0.54 / (0.43) & 0.48 {[}0.44, 0.53{]} & 3 \\
Isosteric & 18 & Net van der Waals (Rosetta, site) \cite{alford2017} &
0.45 / 0.55 / 0.47 / 0.39 & 0.46 {[}0.39, 0.52{]} & 4 \\
Isosteric & 19 & H-bond, side chain (Rosetta, site) \cite{alford2017} &
0.47 / 0.42 / 0.43 / 0.42 & 0.43 {[}0.35, 0.52{]} & 4 \\
Isosteric & 20 & Site total (Rosetta) \cite{alford2017} &
0.41 / 0.58 / 0.41 / 0.29 & 0.42 {[}0.37, 0.47{]} & 4 \\
Isosteric & 21 & Pairwise distance, mean & 0.43 / 0.49 / 0.32 / -- & 0.41 {[}0.34, 0.50{]} & 3 \\
Isosteric & 22 & pKa, mean (PROPKA) \cite{olsson2011,sondergaard2011} &
0.40 / (0.47) / -- / -- & 0.40 {[}0.28, 0.54{]} & 1 \\
Isosteric & 23 & H-bond, backbone to side chain (Rosetta, site) \cite{alford2017} &
0.26 / 0.29 / 0.31 / 0.29 & 0.29 {[}0.19, 0.39{]} & 4 \\
Isosteric & 24 & Electrostatics (Rosetta, site) \cite{alford2017} &
0.26 / 0.20 / 0.24 / 0.26 & 0.24 {[}0.18, 0.30{]} & 4 \\
Isosteric & -- & Polar contacts & (0.93) / (0.97) / (0.96) / (0.96) & -- & 0 \\
Non-isosteric & 1 & AME RMSD \cite{ahern2026} &
0.61 / 0.73 / 0.68 / 0.72 & \textbf{0.68 {[}0.62, 0.75{]}} & 4 \\
Non-isosteric & 2 & lk-ball solvation (Rosetta, site) \cite{alford2017} &
(0.56) / 0.53 / 0.59 / 0.65 & 0.59 {[}0.51, 0.67{]} & 3 \\
Non-isosteric & 3 & Total SASA & (0.46) / 0.51 / (0.49) / 0.58 & 0.55 {[}0.48, 0.60{]} & 2 \\
Non-isosteric & 4 & pKa shift, max absolute (PROPKA) \cite{olsson2011,sondergaard2011} &
(0.53) / 0.53 / -- / -- & 0.53 {[}0.42, 0.65{]} & 1 \\
Ala & 1 & Fraction buried & (0.54) / (0.56) / 0.70 / 0.63 & \textbf{0.66 {[}0.59, 0.74{]}} & 2 \\
Gly & 1 & Omega torsion (Rosetta, site) \cite{alford2017} &
(0.46) / -- / (0.58) / 0.52 & 0.52 {[}0.45, 0.60{]} & 1 \\
Never & 1 & Ligand clearance \cite{ahern2026} &
0.57 / 0.63 / 0.57 / 0.63 & 0.60 {[}0.54, 0.66{]} & 4 \\
Never & -- & Pairwise distance, max & (0.30) / (0.36) / (0.19) / -- & -- & 0 \\
\end{longtable}

\textbf{More detail.}

\begin{itemize}
\tightlist
\item \emph{Ranking.} All 41 quantities scored for the catalytic lesion, the 33 ranked metrics and 8
other rows (6 supplementary metrics and 2 aliases of ranked metrics), with R and the response,
control-noise and specificity flags at each step, and the mean R of the 143 enzymes of the main set
alone: Table S19. The numbers of eligible, responding and specific structure-space metrics per
family and step, with the panel statistic and the four sensitivity sets: Table S20.
\item \emph{Specificity ratio.} The ratio, the interval and the verdict of each eligible metric at
each step: Table S23; the cells that pass the controls: Table S24; R per step on the 143 enzymes of
the main set with 90\% intervals: Table S25; the robustness to motif size, packing matching and
count gating: Table S21; the panel statistic: Table S26; the effective number of independent
metrics: Table S29.
\item \emph{Controls and floors.} The loosest matching level needed and the balance of the residues
changed in both arms: Tables S17 and S18; the burial of the constructed control in the de novo
designs: Table S30; the sanity floors: Table S16.
\item \emph{Predictor ladder.} Each metric at each step: Table S27; the seed noise: Table S22;
PLACER ensemble metrics at the isosteric step: Table S31.
\end{itemize}

\subsubsection{3.4.2 Detection of the second-shell lesion}

\textbf{Experiment.} One, two or four second-shell residues around the catalytic residues are
mutated and the structure is repacked (second-shell removal); the control does the same for the same
number of residues around matched non-catalytic positions, and both arms are scored over the
catalytic residues.

\textbf{Dataset.} The 143 M-CSA enzymes \cite{ribeiro2018} pooled with the 52 enzymes of the
53-enzyme set that are not in it, with a control arm for at most 193 of them; the number of enzymes
in which a metric moves at all ranges from 0 to 182 per metric and dose. Structure-space metrics are
scored on this lesion (Methods 2.5.2).

\textbf{Ranking criteria.} The 29 structure-space metrics of the main set are scored as in 3.4.1,
with R at the three doses (1, 2 and 4 removed residues) and their mean over the doses. A dose only
counts for a metric if the metric responds to the removal (the interval of its median change
excludes zero) and the control response lies above the metric's noise; a dose with fewer than five
informative enzymes has no R. For no metric does a dose qualify, so raw values are reported and no
best-of-M threshold is used (Methods 2.5.2). The 37 quantities scored, with R and its interval at
each dose, the mean R, and the flags of response and of control response above noise at each dose,
are in Table S28.

\textbf{Result.} No metric has a dose at which it both responds to the removal and has a control
response above its noise. The mean pKa shift and the mean pKa respond at 4 removed residues and have
no control response above their noise; 11 other metrics have a control response above their noise
without responding, at one dose (8 metrics) or at all three (the Ramachandran term, the omega
torsion and the amino-acid propensity). The raw mean R of 26 metrics lies between 0.61 and 0.78 (R
at 1 removed residue 0.56 to 0.91, at 2 residues 0.59 to 0.79, at 4 residues 0.57 to 0.69), of which
the largest is that of the fraction buried (0.78 {[}0.68, 0.87{]}). The Ramachandran term has an R of
0.10, 0.00 and 0.00 (mean 0.03 {[}0.00, 0.11{]}), so that at every dose it changed more at the
matched control than at the second-shell removal around the catalytic residues. The omega torsion
and the amino-acid propensity have fewer than five informative enzymes at every dose and no R.

\textbf{More detail.} The 37 quantities scored for the second-shell lesion, the 29 ranked metrics
and 8 other rows (6 supplementary metrics and 2 aliases of ranked metrics), with R and its interval,
the response and control-noise flags at each dose, grouped by the dose at which the metric is first
detected: Table S28. The specificity ratio and the verdict at each dose: Table S23. The numbers of
eligible, responding and specific metrics per family and dose: Table S20.

\begin{center}\rule{0.5\linewidth}{0.5pt}\end{center}

\section{4 Discussion}

Because generative models produce more designs than any laboratory can test, the in-silico filter
that chooses among them has become a decisive step of enzyme design. The designs that are built are
those that pass the filter, and the same metrics increasingly decide which design methods are
judged to be better. A filter that does not track catalysis therefore wastes experiments and can
mislead the field about what works, and yet, to our knowledge, the metrics behind these filters had
not yet been tested against catalysis.

We tested 33 such metrics in four tests, each corresponding to a task that is asked of a filter.
Activity discrimination is the basic one: telling designs that work from designs that do not.
Substrate discrimination asks whether a metric knows what an enzyme was built for, which a design
conditioned on a substrate takes for granted. Activity ranking is the last step of a selection:
choosing the best of the designs that pass. Detection ability asks for the property that makes a
metric a measure of catalysis and not of geometry: it must register damage to the catalytic residues
more strongly than the same damage elsewhere in the protein. Three of the tests observe proteins as
they are and the fourth intervenes, which makes it possible to compare the response at the
catalytic residues with the response at a matched control. A metric that tracks catalysis should
score well in all four.

In activity discrimination, the intra-residue repulsion of the catalytic residues ranked first on
natural dead and active enzymes, with an AUROC of 0.61 against a chance threshold of 0.68, and on
192 designs (0.80), ahead of the repulsion of the catalytic residues (0.78). In substrate
discrimination, ipTM ranked first (0.70) and the RMSD of the catalytic atoms reached 0.53 to 0.56.
In activity ranking, the catalytic pKa range correlated best with the measured impairment of the
single-point variants of the enzyme BglB (ρ = 0.48), a task that differs from ranking de novo
designs. In detection ability, the per-chain pTM minimum had the highest mean R over the counted
steps (0.75). The 192 designs, of which 16 have a reported kcat/KM, had already passed the filters
of the campaign that produced them, and the remaining 176 have no reported kcat/KM.

\section{5 Data availability}

The Supplement gives the full per-metric tables behind Tables 5--10 and further analyses. All tables
are released as CSV files at \url{https://github.com/RIA-lab/MetricResults} (tables of the main text
in \texttt{main/}, supplementary tables in \texttt{supplementary/}); the Supplement PDF describes
each table and gives the path of the corresponding file. Structures and weights of the predictor are
third-party and are not redistributed again. BglB data at variant level are not redistributed
because the licence is not stated; only aggregates are reported.

\bibliographystyle{unsrt}
\bibliography{references}

\end{document}